\documentclass[10pt]{article}

\usepackage{amsmath,amssymb,amsthm,bm}

\usepackage{algorithmic}
\usepackage[]{algorithm}

\usepackage[pdftex]{graphicx}

\usepackage{cite}

\usepackage{color}
\usepackage{geometry}  
\usepackage{setspace}
\usepackage[table]{xcolor}
\usepackage[labelfont=bf,labelsep=period,justification=raggedright]{caption}

\makeatletter
\renewcommand{\@biblabel}[1]{\quad#1.}
\makeatother

\date{}

 \definecolor{lblue}{rgb}{0.8,0.9,1}

\begin{document}

% Title must be 150 characters or less
\begin{flushleft}
{\Large
\textbf{Canalization and  control in automata networks: \\body segmentation in \emph{Drosophila melanogaster}}
}
% Insert Author names, affiliations and corresponding author email.
\\
Manuel Marques-Pita$^{1,2}$,
Luis M. Rocha$^{2,1}$,
\\\bf{1} Instituto Gulbenkian de Ci\^encia. Oeiras. Portugal
\\
\bf{2} Indiana University. Bloomington, IN. USA
\\
E-mail: marquesm@indiana.edu, rocha@indiana.edu
\end{flushleft}

% Please keep the abstract between 250 and 300 words
\section*{Abstract}
%\addcontentsline{toc}{section}{Abstract}

We present schema redescription as a methodology to
characterize canalization in automata networks used to model
biochemical regulation and signalling.
In our formulation, canalization becomes synonymous with redundancy
present in the logic of automata.
This results in straightforward measures to quantify canalization in
an automaton (micro-level), which is in turn integrated into a highly
scalable framework to characterize the collective dynamics of large-scale
automata networks (macro-level).
This way, our approach provides a method to link micro- to macro-level
dynamics -- a crux of complexity.
Several new results ensue from this methodology: uncovering of
dynamical modularity (modules in the dynamics rather than in the
structure of networks), identification of minimal conditions and
critical nodes to control the convergence to attractors, simulation of
dynamical behaviour from incomplete information about initial
conditions, and measures of macro-level canalization and robustness to
perturbations.
We exemplify our methodology with a well-known model of the
intra- and inter cellular genetic regulation of body segmentation in
\emph{Drosophila melanogaster}.
We use this model to show that our analysis does not contradict any
previous findings. But we also obtain new knowledge about its
behaviour: a better understanding of the size of its wild-type
attractor basin (larger than previously thought), the
identification of novel minimal conditions and critical nodes that
control wild-type behaviour, and the resilience of these to
stochastic interventions.
Our methodology is applicable to any complex network that can be
modelled using automata, but we focus on biochemical
regulation and signalling, towards a better understanding of the
(decentralized) control that orchestrates cellular activity -- with
the ultimate goal of explaining how do cells and tissues `compute'.

% Please keep the Author Summary between 150 and 200 words
% Use first person. PLoS ONE authors please skip this step.
% Author Summary not valid for PLoS ONE submissions.
%\section*{Author Summary}

\section*{Introduction and background}
%\addcontentsline{toc}{section}{Introduction and background}

The notion of \emph{canalization} was proposed by Conrad Waddington
\cite{Waddington:1942sd} to explain why, under genetic and
environmental perturbations, a wild-type phenotype is less variable in
appearance than most mutant phenotypes during development.
Waddington's fundamental hypothesis was that the robustness of
wild-type phenotypes is the result of a \emph{buffering of the
  developmental process}.
This led Waddington to develop the well-known concept of
\emph{epigenetic landscape} \cite{Waddington:1957pi,conrad1990}, where
cellular phenotypes are seen, metaphorically, as marbles rolling down
a sloped and ridged landscape as the result of interactions amongst
genes and epigenetic factors.
The marbles ultimately settle in one of the valleys, each
corresponding to a stable configuration that can be reached via the
dynamics of the interaction network.
In this view, genetic and epigenetic perturbations can only have a
significant developmental effect if they force the natural path of the
marbles over the ridges of the epigenetic landscape, thus making them
settle in a different valley or stable configuration.

Canalization, understood as the buffering of genetic and epigenetic
perturbations leading to the stability of phenotypic traits, has
re-emerged recently as a topic of interest in computational and
systems biology
\cite{Fraser:2010ly,Tusscher:2009ys,Gibbon:2009hc,Levy:2008ve,Masel:2007zr,Bergman:2003qf,Siegal:2002lq}.
However, canalization is an emergent phenomenon because we can
consider the stability of a phenotypic trait both at the micro-level
of biochemical interactions, or at the macro-level of phenotypic
behaviour.
The complex interaction between micro- and macro-level thus makes
canalization difficult to study in biological organisms -- but  the
field of complex systems has led to progress in our understanding of
this concept.
For instance, Conrad \cite{conrad1990} provided a still-relevant
treatment of evolvability \cite{pigliucci2008} by analysing the
tradeoff between genetic (micro-level) instability and phenotypic
(macro-level) stability. 
This led to the concept of \emph{extra-dimensional bypass}, whereby
most genetic perturbations are buffered to allow the phenotype to be
robust to most physiological perturbations, but a few genetic
perturbations (e.g. the addition of novel genetic information) provide
occasional instability necessary for evolution.
Conrad highlighted three (micro-level) features of the organization of
living systems that allows them to satisfy this tradeoff:
\emph{modularity} (or compartmentalization), \emph{component
  redundancy}, and \emph{multiple weak interactions}.
The latter two features are both a form of redundancy, the first
considering the redundancy of components and the second considering
the redundancy of interactions or linkages.
Perhaps because micro-level redundancy has been posited as one of the
main mechanisms to obtain macro-level robustness, the term
canalization has also been used -- especially in discrete
mathematics -- to characterize redundant properties of automata
functions, particularly when used to model micro-level dynamical
interactions in models of genetic regulation and biochemical
signalling.

An automaton is typically defined as \emph{canalizing} if there is at
least one state of at least one of its inputs that is
sufficient to control the automaton's next state (henceforth
\emph{transition}), regardless of the states of any other inputs
\cite{Kauffman:1984uq}.
Clearly, this widely used definition refers to micro-level
characteristics of the components of multivariate discrete dynamical
systems such as automata networks, and not to canalization as the
emergent phenomenon outlined above.
Nonetheless, using this definition, it has been shown that
(1) canalizing functions are widespread in eukaryotic gene-regulation
dynamics \cite{Harris:2002qa};
(2) genetic regulatory networks modelled with canalizing automata are
always stable \cite{Kauffman:2003ys,Kauffman:2004sy};
and (3) realistic biological dynamics are naturally observed in networks with
scale-free connectivity that contain canalizing functions
\cite{Grefenstette:2006rv}.
These observations suggest that the redundancy captured by this
micro-level definition of canalization is a mechanism used to obtain
stability and robustness at the macro-level of phenotypic traits.

Since the proportion of such `strictly' canalizing functions drops
abruptly with the number of inputs ($k$) \cite{Raeymaekers:2002zh}, it
was at first assumed that (micro-level) canalization does not play a
prominent role in stabilizing complex dynamics of gene regulatory
networks.
However, when the concept of canalization is extended to include
\emph{partially canalizing} functions, where subsets of inputs can
control the automaton's transition, the proportion of available
canalizing automata increases dramatically even for automata with many
inputs \cite{Reichhardt:2007bs}.
Furthermore, partial canalization has been shown to contribute to network
stability, without a detrimental effect on `evolvability' \cite{Reichhardt:2007bs}.
Reichhardt and Bassler, point out that, even though strictly
canalizing functions clearly contribute to network stability, they can
also have a detrimental effect on the ability of networks to adapt to
changing conditions \cite{Reichhardt:2007bs} -- echoing Conrad's
tradeoff outlined above.
This led them to consider the wider class of partially canalizing
functions that confer stable network dynamics, while improving
adaptability.
A function of this class may ignore one or more of its inputs given
the states of others, but is not required to have a single canalizing input.
For example, if a particular input is \emph{on}, the states of
the remaining inputs are irrelevant, but if that same input is \emph{off},
then the state of a subset of its other inputs is required
to determine the function's transition.
In scenarios where two or more inputs are needed to determine the
transition, the needed inputs are said to be \emph{collectively
  canalizing}.

Reichhardt and Bassler \cite{Reichhardt:2007bs} have shown that
the more general class of partially canalizing functions dominates the
space of Boolean functions for any number of inputs $k$.
Indeed, for any value of $k$, there are only two \emph{non-canalizing
  functions} that always depend on the states of all inputs.
Other classes of canalizing functions have been considered, such as
\emph{nested canalizing} functions  \cite{Kauffman:2003ys},
\emph{Post classes} \cite{Shmulevich:2003kl} and \emph{chain
  functions} \cite{Gat-Viks:2003ge}.
All these classes of functions characterize situations of input
redundancy in automata. 
In other words, micro-level canalization is understood as a form of
redundancy, whereby a subset of input states is sufficient to
guarantee transition, and therefore its complement subset of input
states is redundant.
This does not mean that redundancy is necessarily the sole -- or even
most basic -- mechanism to explain canalization at the macro-level.
But the evidence we reviewed above, at the very least, strongly
suggests that micro-level redundancy is a key mechanism to achieve
macro-level canalization.
Other mechanisms are surely at play, such as the topological
properties of the networks of micro-level interactions. 
Certainly, modularity, as suggested by Conrad, plays a role in the
robustness of complex systems and has rightly received much attention
recently \cite{Fortunato:2009ly}.
While we show below that some types of modularity can derive from
micro-level redundancy, other mechanisms to achieve modularity are
well-known \cite{Fortunato:2009ly}.

Here, we explore partial canalization, as proposed by Reichhardt and
Bassler\cite{Reichhardt:2007bs}, to uncover the loci of control in
complex automata networks, particularly those used as models of
genetic regulation and signalling.
Moreover, we extend this notion to consider not only (micro-level)
canalization in terms of input redundancy, but also in terms of
input-permutation redundancy to account for the situations in which
swapping the states of (a subset) of inputs has no effect on an
automaton's transition.
From this point forward, when we use the term \emph{canalization} we
mean it in the micro-level sense used in the (discrete dynamical
systems) literature to characterize redundancy in automata functions.
Nonetheless, we show that the quantification of such micro-level
redundancy uncovers important details of macro-level dynamics in
automata networks used to model biochemical regulation.
This allows us to better study how robustness and control of
phenotypic traits arises in such systems, thus moving us towards
understanding canalization in the wider sense proposed by Waddington.
Before describing our methodology, we introduce necessary concepts and
notations pertaining to Boolean automata and networks, as well as the
segment polarity gene-regulation network in \emph{Drosophila
  melanogaster}, an automata model we use to exemplify our approach.

\subsection*{Boolean networks}
%\addcontentsline{toc}{subsection}{Boolean networks}

\begin{figure}
\includegraphics{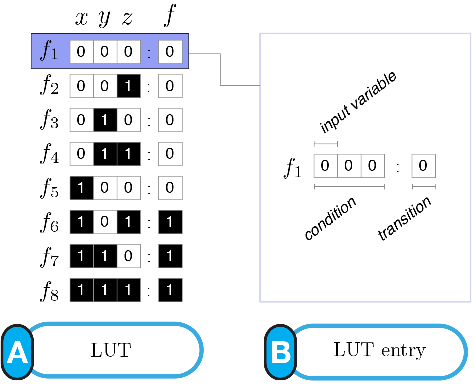}
\caption{{\bf (A) LUT for Boolean automaton $f = x \wedge
(y \vee z)$ and (B) components of a single LUT entry}.}
\label{fig:lut_example}
\end{figure}

This type of discrete dynamical system was introduced to build
qualitative models of genetic regulation, very amenable to large-scale
statistical analysis \cite{Kauffman:1969fk} -- see
\cite{Gershenson:2004bh} for comprehensive review.
A \emph{Boolean automaton} is a binary variable, $x \in
\{0,1\}$, where state 0 is interpreted as \emph{false} (\emph{off} or \emph{unexpressed}), and
state 1 as \emph{true} (\emph{on} or \emph{expressed}).
The states of $x$ are updated in discrete time-steps, $t$,
according to a \emph{Boolean state-transition function} of $k$ inputs:
$x^{t+1} = f\left(i_1^t, ..., i_k^t\right)$.
Therefore $f: \{0,1\}^k \rightarrow \{0,1\}$.
Such a function can be defined by a \emph{Boolean logic formula} or by
a \emph{look-up (truth) table} (LUT) with $2^{k}$ entries.
An example of the former is $x^{t+1} = f(x,y,z) = x^t \wedge (y^t \vee
z^t)$, or its more convenient shorthand representation $f = x \wedge
(y \vee z)$, which is a Boolean function of $k=3$ input binary
variables $x,y,z$, possibly the states of other automata; $\wedge$,
$\vee$ and $\neg$ denote logical conjunction, disjunction, and
negation respectively. The LUT for this function is shown in Figure
\ref{fig:lut_example}.
Each LUT entry of an automaton $x$, $f_{\alpha}$, is defined by (1) a
specific
\emph{condition}, which is a conjunction of $k$ inputs represented as
a unique $k$-tuple of input-variable (Boolean) states, and (2) the
automaton's \emph{next state} (transition) $x^{t+1}$, given the condition (see
Figure \ref{fig:lut_example}).
We denote the entire state transition function of an automaton $x$ in its
LUT representation as $F \equiv \{f_\alpha: \alpha = 1,...,2^k\}$.

A \emph{Boolean Network} (BN) is a graph $\mathcal{B} \equiv (X,E)$,
where $X$ is a set of $n$ Boolean automata \emph{nodes}
$x_i \in X, i=1,...,n$, and $E$ is a set of directed edges $e_{ji} \in
E: x_i,x_j \in X$.
If $e_{ji} \in E$, it means that automaton $x_j$ is an
input to automaton $x_i$, as computed by $F_i$.
$X_i = \{ x_j \in X : e_{ji} \in E\}$ denotes the set of input automata of
$x_i$.
Its cardinality, $k_i = |X_i|$, is the \emph{in-degree} of node $x_i$,
which determines the size of its LUT, $|F_i| = 2^{k{_i}}$.
We refer to each entry of $F_i$ as $f_{i:\alpha}, \alpha = 1...2^{k{_i}}$.
The \emph{input nodes} of $\mathcal{B}$ are nodes whose state does not
depend on the states of other nodes in $\mathcal{B}$.
The state of \emph{output nodes} is determined by the states of other
nodes in the network, but they are not an input to any other
node. Finally, the state of \emph{inner nodes} depends on the state of
other nodes and affect the state of other nodes in $\mathcal{B}$.
At any given time $t$, $\mathcal{B}$ is in a specific
\emph{configuration} of node states, $\bm{x}^t = \langle x_1, x_2,
..., x_n\rangle$.
We use the terms \emph{state} for individual automata $(x)$ and
\emph{configuration} $(\bm{x})$ for the collection of states of the set
of automata of $\mathcal{B}$, i.e. the collective network state.

Starting from an initial configuration, $\bm{x}^0$, a BN updates its
nodes with a \emph{synchronous}
or \emph{asynchronous} policy.
The \emph{dynamics} of $\mathcal{B}$ is thus defined by
the temporal sequence of configurations that ensue, and there are
$2^n$ possible configurations.
The transitions between configurations  can be
represented as a \emph{state-transition graph}, $\textrm{STG}$,
where each vertex is a configuration, and each directed edge denotes a
transition from $\bm{x}^t$ to $\bm{x}^{t+1}$.
The STG of $\mathcal{B}$ thus encodes the network's entire
\emph{dynamical landscape}.
Under synchronous updating, configurations that repeat, such that
$\bm{x}^{t+\mu} = \bm{x}^t$, are known as \emph{attractors};
\emph{fixed point} when $\mu=1$, and \emph{limit cycle} -- with period
$\mu$ -- when $\mu > 1$, respectively.
The disconnected subgraphs of a STG leading to an
attractor are known as \emph{basins of attraction}.
In contrast, under asynchronous updating, there are alternative
configuration transitions that depend on the order in which nodes
update their state.
Therefore, the same initial configuration can converge to distinct
attractors with some probability \cite{Thomas:1973fk,Thomas:1995qf}.
A BN $\mathcal{B}$ has a finite number $b$ of attractors; each
denoted by $\mathcal{A}_i  :  i=1,...,b$.
When the updating scheme is known, every configuration $\bm{x} $ is
in the basin of attraction of some specific attractor $\mathcal{A}_i$.
That is, the dynamic trajectory of $\bm{x}$ converges to
$\mathcal{A}_i$.
We denote such a dynamical trajectory by $\sigma(\bm{x}) \leadsto
\mathcal{A}_i$.
If the updating scheme is stochastic, the relationship between
configurations and attractors can be specified as the conditional
probability $P(\mathcal{A}_i | \bm{x})$.

\subsection*{The segment polarity network}
%\addcontentsline{toc}{subsection}{SPN}

The methodology introduced in this paper will be exemplified using the
well-studied Boolean model of the segment polarity network in
\emph{Drosophila melanogaster} \cite{Albert:2003ij}.
During the early ontogenesis of the fruit fly, like in every
arthropod's development, there is body segmentation
\cite{Alberts:2003xw, Wolpert:1998ee}.
The specification of adult cell types in each of these segments is
controlled by a hierarchy of around forty genes.
While the effect of most of the genes in the hierarchy is only
transient, a subset of \emph{segment polarity genes} remains expressed
during the life of the fruit fly \cite{Hooper:1992zl}.
The dynamics of the segment polarity network was originally modelled using a
system of non-linear ordinary differential equations (ODEs)
\cite{Dassow:2000qd, Dassow:2002rc}.
This model suggested that the regulatory network of segment polarity
genes is a module largely controlled by external inputs that is robust
to changes to its internal kinetic parameters.
On that basis, Albert and Othmer later proposed a simpler discrete BN
model of the dynamics of the \emph{segment polarity network}
(henceforth SPN) \cite{Albert:2003ij}.
This was the first Boolean model of gene regulation capable of
predicting the steady state patterns experimentally observed in wild-type
and mutant embryonic development with significant accuracy, and has
thus become the quintessential example of the power of the logical
approach to modelling of biochemical regulation from qualitative data in the literature.
Modelling with ODEs, in contrast, is hindered by the need of substantial
quantitative data for parameter estimation
\cite{Aldana:2007oq,Bornholdt:2008kl,Irons:2009tg,Alvarez-Buylla:2008fk,Samaga:2010bs,Assmann:2009dz}.

\begin{figure}
\includegraphics{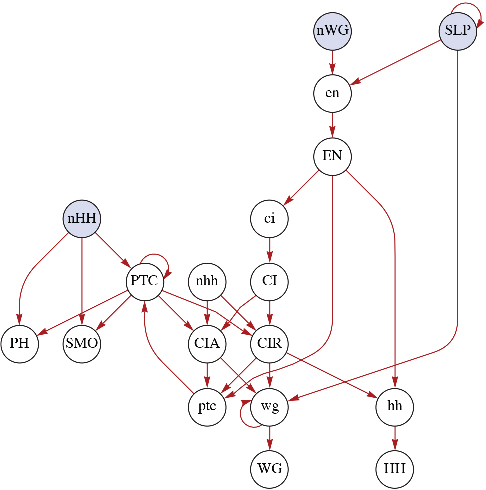}
\caption{{\bf Connectivity graph of the SPN model}.  The fifteen
  genes and proteins considered in the SPN model are represented
  (white nodes). The incoming edges to a node $x$ originate in the
  nodes that are used by $x$ to determine its transition. Shaded nodes
  represent the spatial signals (states of WG, HH and $hh$ in
  neighbouring cells). Note that SLP -- derived from an upstream
  intra-cellular signal -- is an \emph{input} node to this
  network. The self-connection it has represents the steady-state
  assumption: $SLP_i^{t+1} = SLP_i^t$. Notice also that this graph
  represents the fully synchronous version of the model, where
  modifications concerning PH and SMO have been made (see main text
  for details).}
\label{fig:spn_bn}
\end{figure}

\begin{figure}
\includegraphics{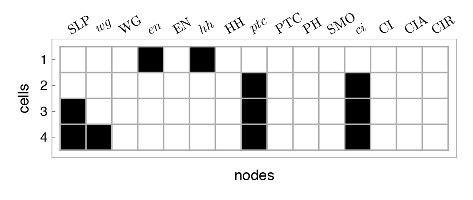}
\caption{{\bf A parasegment in the SPN model}. Cells are represented
  horizontally, where the top (bottom) row is the most anterior
  (posterior) cell. Each column is a gene, protein or complex -- a node
  in the context of the BN model. The specific pattern shown
  corresponds to the wild-type initial expression pattern observed at
  the onset of the segment polarity genes regulatory dynamics
  ($\bm{x}_{\textrm{ini}}$); Black/on (white/off) squares represent
  expressed (not expressed) genes or proteins.}
\label{fig:spn_parasegment}
\end{figure}

The SPN model comprises fifteen nodes that represent intra-cellular
chemical species and the genes \emph{engrailed (en)};
\emph{wingless (wg)}; \emph{hedgehog (hh)}; \emph{patched (ptc)} and
\emph{cubitus interruptus (ci)} \cite{Hooper:1992zl,Dassow:2000qd,
  Dassow:2002rc}.
These genes encode a number of proteins such as the transcription
factors Engrailed (EN), Cubitus Interruptus (CI), CI Activator (CIA),
and CI repressor (CIR); the secreted proteins Wingless (WG) and
Hedgehog (HH); and the transmembrane protein Patched (PTC).
Other proteins included in the SPN model are Sloppy-Paired (SLP) --
the state of which is previously determined by the \emph{pair-rule} gene family
that stabilizes its expression before the segment polarity genes -- as
well as Smoothened (SMO) and the PH complex that forms when HH from
neighbouring cells binds to PTC.
Figure \ref{fig:spn_bn} shows the topology and Table
\ref{tab:SPN_Logic_Table} lists the logical rules of the nodes in every cell of the
SPN.
This model consists of a spatial arrangement of four interconnected
cells, a \emph{parasegment}.
While the regulatory interactions within each cell are governed by the
same network, inter-cellular signalling affects neighbouring
cells.  That is, regulatory interactions in a given cell depend on the states of WG, $hh$ and HH in adjacent cells.
Therefore, six additional (inter-cellular) `spatial signals' are
included: $hh_{i \pm 1}$, $\text{HH}_{i \pm 1}$ and $\text{WG}_{i
  \pm 1}$, where $i = 1, ... ,4$ is the cell index in the four-cell
parasegment.
In a parasegment, the cell with index $i=1$ corresponds to its
anterior cell and the cell with index $i=4$ to its posterior cell (see
Figure \ref{fig:spn_parasegment}).
In simulations, the four-cell parasegments assume periodic boundary
conditions (i.e. anterior and posterior cells are adjacent to each other).
Since each parasegment has $4 \times 15 = 60$ nodes, four of which
are in a fixed state (SLP), there are $2^{56}$ possible
configurations -- a dynamical landscape too large for exhaustive
analysis.
Even though the original model was not fully synchronous because
$\text{PH}$ and $\text{SMO}$ were updated instantaneously at time
$t$, rather than at $t+1$, here we use the fully equivalent, synchronous
version.
Specifically, since $\text{PH}$ is an output node, synchronizing its
transition with the remaining nodes at $t+1$ does not impact the
model's dynamics.
The state of $\text{SMO}$ influences the updating of $\text{CIA}$ and
$\text{CIR}$;
but since the update of $\text{SMO}$ is instantaneous, we can include
its state-transition function in the state-transition functions of $\text{CIA}$ and
$\text{CIR}$ (which update at $t+1$) with no change in the dynamics of
the model as described in \cite{Chaves:2005fk}.

\begin{figure}
   \includegraphics{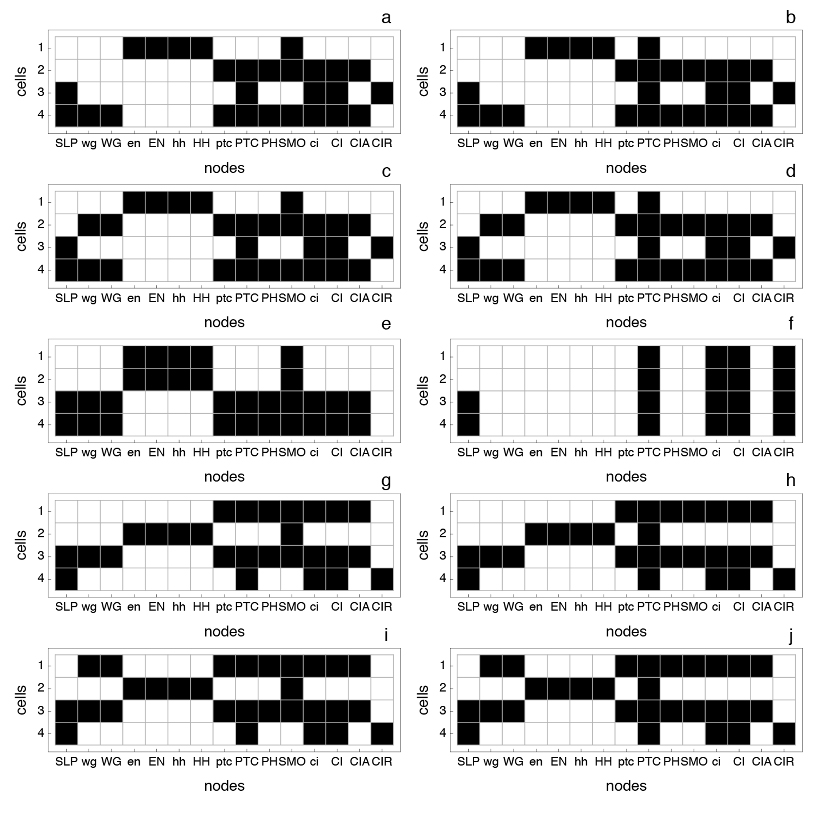}
   \caption{{\bf The ten attractors reached by the SPN.} These
     attractors are divided in four groups: wild-type, broad-stripe,
     no segmentation and ectopic. More specifically: (a) wild-type,
     (b) variant of (a), (c) wild-type with two $wg$ stripes, (d)
     variant of (c), (e) broad-stripe, (f) no segmentation, (g)
     ectopic, (h) variant of (g), (i) ectopic with two $wg$ stripes,
     and (j) variant of (i). The wild-type attractor (a) is
     referred to as $\mathcal{A}_{\textrm{wt}}$ in the results
     and discussion sections. }
\label{fig:spatial_attractors}
\end{figure}

The initial configuration (IC) of the SPN, depicted in Figure
\ref{fig:spn_parasegment}, and which leads to the wild-type expression
pattern is known \cite{Albert:2003ij}: $wg_4 = en_1 = hh_1 =
ptc_{2,3,4} = ci_{2,3,4} = 1$ (\emph{on} or expressed).
The remaining nodes in every cell of the parasegment are set to 0
(\emph{off}, or not expressed).
Overall, the dynamics of the SPN settles to one of ten
attractors -- usually divided into four qualitatively distinct groups,
see Figure \ref{fig:spatial_attractors}:
(1) wild-type with three extra variations (PTC mutant, double $wg$
bands, double $wg$ bands PTC mutant); (2) Broad-stripe mutant; (3) No
segmentation; and (4) Ectopic (with the same variations as wild-type).
Albert and Othmer estimated that the number of configurations that
converge to the wild-type attractor is approximately $6\times10^{11}$
-- a very small portion of the total number of possible
configurations ($\approx 7 \times 10^{16}$) -- and that the broad-stripe mutant
attractor basin contains about $90\%$ of all possible configurations
\cite{Albert:2003ij}.

The inner and output nodes of each cell in a parasegment -- that is,
every node except the input node SLP -- that has reached a stable
configuration (attractor) are always in one of the following five
patterns.

\begin{itemize}

\item{$I1$}: all nodes are \emph{off} except PTC, $ci$, CI and CIR.
\item{$I2$}: same as $I1$ but states of $ptc$, PH, SMO, CIA and CIR are negated.
\item{$I3$}: all nodes are \emph{off} except $en,$ EN, $hh$, HH and SMO.
\item{$I4$}: same as $I3$ but PTC and SMO are negated.
\item{$I5$}: negation of $I4$, except PTC and CIR remain as in $I4$.
\end{itemize}

\noindent For example, the wild-type configuration corresponds -- from
anterior to posterior cell -- to the patterns $I3$, $I2$, $I1$ and
$I5$.
Thus the pattern $I4$ is only seen in mutant expression patterns.
The patterns $I1$ to $I5$ can be seen as attractors of the inner- and
output-node dynamics of each cell in a parasegment.

Besides the fact that the SPN is probably the most well-known discrete
dynamical system model of biochemical regulation, we chose it to
exemplify our methodology because
(1) it has been well-validated experimentally, despite the assumption that genes and proteins operate like
\emph{on/off} switches with synchronous transitions
and (2) the model includes both intra-cellular regulation and
inter-cellular signalling in a spatial array of cells.
The intra and inter-cellular interactions in the SPN model result in
a dynamical landscape that is too large to characterize via an STG,
while adding also an extra level of inter-cellular (spatial) regulation.
The ability to deal with such multi-level complexity makes our
methodology particularly useful.
As we show below, we can uncover the signals that control collective
information processing in such (spatial and non-spatial) complex
dynamics.

\section*{Methodology and Results}
%\addcontentsline{toc}{section}{Methodology and results}

\subsection*{Micro-level canalization via schemata}
%\addcontentsline{toc}{subsection}{Micro-level canalization}

In previous work, we used \emph{schema redescription} to demonstrate
that we can understand more about the dynamical behaviour of automata
networks by analysing the patterns of \emph{redundancy} present in their (automata)
components (micro-level), rather than looking solely at their
macro-level or collective behaviour \cite{Marques-Pita:2011uq}.
Here we relate the redundancy removed via schema redescription with
the concept of \emph{canalization}, and demonstrate that a
characterization of the full canalization present in biochemical
networks leads to a better understanding of how cells and collections of
cells `compute'.
Moreover, we show that this leads to a comprehensive characterization of
\emph{control} in automata models of biochemical regulation.
Let us start by describing the schema redescription methodology.
Since a significant number of new concepts and notations are
introduced in this, and subsequent sections, a succinct glossary of
terms as well as a table with the mathematical notations used is
available in \emph{Supporting text S1}.

From the extended view of canalization introduced earlier, it follows
that the inputs of a given Boolean automaton do not control its
transitions equally.
Indeed, substantial redundancy in state-transition functions is
expected.
Therefore, filtering redundancy out is equivalent to
identifying the loci of control in automata.
In this section we focus on identifying the loci of control in
individual automata by characterizing the canalization present in
their transition functions.
First, we consider how subsets of inputs in specific state
combinations make other inputs \emph{redundant}.
Then we propose an additional form of canalization that accounts for
\emph{input permutations} that leave a transition unchanged.
Later, we use this characterization of canalization and control in
individual automata to study networks of automata; this also allows us
to analyse robustness and collective computation in these networks.

\subsubsection*{Wildcard schemata and \emph{enputs}}
%\addcontentsline{toc}{subsubsection}{Wildcard schemata}

\begin{figure}
\includegraphics{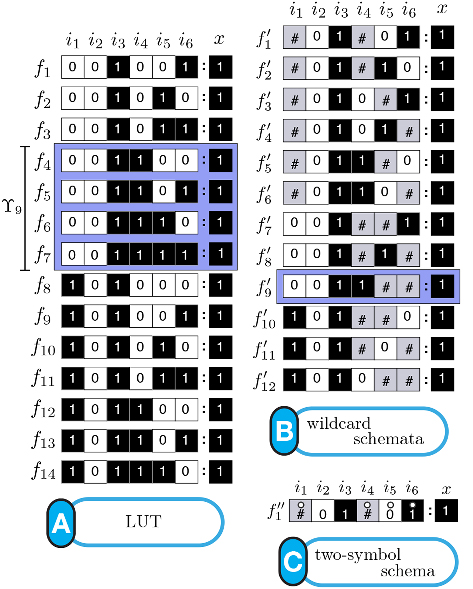}
\caption{{\bf Schema redescription.} (A) Subset of LUT entries of an
  example automaton $x$ that prescribe state transitions to \emph{on}
  (1); white (black) states are 0 (1). (B) Wildcard schema
  redescription; wildcards denoted also by grey states.  Schema $f'_9$
  is highlighted to identify the subset of LUT entries $\Upsilon_9
  \equiv \{f_4,f_5,f_6,f_7\}$ it redescribes.  (C) Two-symbol schema
  redescription, using the additional position-free symbol; the entire
  set $F'$ is compressed into a single two-symbol schema: $f''_1$. Any
  permutation of the inputs marked with the position-free symbol in
  $f''_1$ results in a schema in $F'$. Note that $f''_1$ redescribes
  the entire set of entries with transition to \emph{on} and thus
  $|\Theta_\theta|=14$. Since there is only one set of marked inputs,
  the position-free symbol does not require an index. Although this
  figure depicts only the schemata obtained for the subset of LUT
  entries of $x$ that transition to \emph{on}, entries that do not
  match any of these schemata transition to \emph{off} (since $x$ is a
  Boolean automaton).}
\label{fig:schema_example}
\end{figure}

Consider the  example automaton $x$ in Figure \ref{fig:schema_example}A,
where the entire subset of LUT entries in $F$ with transitions to
\emph{on} is depicted.
This portion of entries in $F$ can be \emph{redescribed} as a set of
\emph{wildcard schemata}, $F' \equiv \{f'_{\upsilon}\}$.
A wildcard schema $f'_{\upsilon}$ is exactly like a LUT entry, but
allows an additional \emph{wildcard} symbol, $\#$ (also represented
graphically in grey), to appear in its condition (see Figure
\ref{fig:schema_example}B).
A wildcard input means that it \emph{accepts any state, leaving the
  transition unchanged}.
In other words, wildcard inputs are \emph{redundant} given the
non-wildcard input states specified in a schema $f'_{\upsilon}$.
More formally, when the truth value of an input Boolean variable $i$ in a schema
$f'_{\upsilon}$ is defined by the third (wildcard) symbol, it
is equivalent to stating that the truth value of automaton $x$ is
unaffected by the truth value of $i$ given the conditions defined by $f'_{\upsilon}$: $(x | f'_{\upsilon}, i) = (x
| f'_{\upsilon}, \neg i)$.
Each schema redescribes a subset of entries in the original LUT,
denoted by $\Upsilon_{\upsilon} \equiv \{f_{\alpha}: f_{\alpha}
\rightarrowtail f'_{\upsilon}\}$ ($\rightarrowtail$ means `is
redescribed by').%

Wildcard schemata are \emph{minimal} in the sense that none of the
(non-wildcard) inputs in the condition of a schema can be
`raised' to the wildcard status and still ensure the automaton's
transition to the same state.
Because wildcard schemata are minimal, $\Upsilon_{\upsilon}
\nsubseteq \Upsilon_{\phi} \wedge \Upsilon_{\phi} \nsubseteq
\Upsilon_{\upsilon}, \forall f'_\upsilon, f'_\phi \in F'$.
In other words, a wildcard schema is \emph{unique} in the sense that
the subset of LUT entries it redescribes is not fully redescribed by
any other schema.
However, in general $\Upsilon_{\upsilon} \cap \Upsilon_{\phi} \neq
\emptyset$. This means that schemata can overlap in terms of the LUT
entries they describe.
In Figure \ref{fig:schema_example}, $\Upsilon_1
\equiv \{f_1,f_5,f_9,f_{13}\}$ and $\Upsilon_9 \equiv
\{f_4,f_5,f_6,f_7\}$, therefore $\Upsilon_1 \cap \Upsilon_9 \equiv
\{f_5\}$.
The set of wildcard schemata $F'$ is also \emph{complete}.
This means that for a given LUT $F$ there is one and only one set $F'$
that contains all possible minimal and unique wildcard schemata.
Since wildcard schemata are \emph{minimal,} \emph{unique} and they
form a \emph{complete} set $F'$, they are equivalent to the set of
\emph{all prime implicants} obtained during the first step of the
Quine \& McCluskey Boolean minimization algorithm
\cite{Quine:1955ec}.
Typically, prime implicants are computed for the fraction of the LUT
that specifies transitions to \emph{on}. Then a subset of the
so-called \emph{essential} prime implicants is identified. The set of
essential prime implicants is the subset of prime implicants
sufficient to describe (cover) every entry in the input set of LUT
entries. However, to study how to control the transitions of automata
we use the set of all prime implicants, since it encodes every
possible way a transition can take place.
The set $F'$ may also contain any original entry in $F$ that could not be
subsumed by a wildcard schema.
Although the upper bound on the size of $F'$ is known to be
$O(3^k/\sqrt{k})$ \cite{Chandra:1978fk}, the more input redundancy
there is, the smaller the cardinality of $F'$.

The condition of a wildcard schema can always be expressed as a
logical conjunction of literals (logical variables or their negation),
which correspond to its non-wildcard inputs.
Since a wildcard schema is a \emph{prime implicant}, it follows that
every literal is \emph{essential} to determine the automaton's
transition.
Therefore, we refer to the literals in a schema as its
\emph{essential input states}, or \emph{enputs} for short.
To summarize, each enput in a schema is essential, and the conjunction
of its enputs is a sufficient condition to
\emph{control} the automaton's transition.
It also follows that the set $F'$ of wildcard schemata can be
expressed as a \emph{disjunctive normal form} (DNF) -- that is,
a disjunction of conjunctions that specifies the list of sufficient
conditions to control automaton $x$, where each disjunction clause is a schema.
The DNF comprising all the prime implicants of a Boolean function $f$
is known as its \emph{Blake's canonical
  form} \cite{Blake:1938uq}.
The canonical form of $f$ always preserves the
input-output relationships specified by its LUT $F$.
Therefore, the basic laws of Boolean logic -- contradiction, excluded
middle and de Morgan's laws -- are preserved by the schema redescription.

Schema redescription is related to the work of John Holland on
condition/action rules to model inductive reasoning in cognitive
systems \cite{Holland:1986ly} and to the general \emph{RR framework}
proposed by Annette Karmiloff-Smith to explain the emergence of
internal representations and external notations in human cognition
\cite{Karmiloff93}.
Our methodology to remove redundancy from automata LUTs also bears
similarities with the more general \emph{mask analysis} developed by
George Klir in his `reconstructability' analysis, which is applicable to
any type of variable \cite{Klir:2002uq}.
In addition, prime implicants have been known and used for the
minimization of circuits in electrical engineering since the notion
was introduced by Quine \& McCluskey \cite{Quine:1955ec}; similar
ideas were also used by Valiant \cite{Valiant:1984fk} when introducing
\emph{Probably Approximately Correct} (PAC) learning.

\subsubsection*{Two-symbol schemata}
%\addcontentsline{toc}{subsubsection}{Two-symbol schemata}

We now introduce a different and complementary form of redundancy in
automata, which we consider another form of canalization.
Wildcard schemata identify input states that are sufficient for controlling
an automaton's transition (enputs).
Now we identify subsets of inputs that can be permuted in a schema
without effect on the transition it defines
\cite{Marques-Pita:2011uq}.
For this, a further redescription process takes as input the set of
wildcard schemata ($F'$) of $x$  to compute a set of
two-symbol schemata $F'' \equiv \{f''_{\theta}\}$ (see Figure
\ref{fig:schema_example}C).
The additional \emph{position-free symbol} ($\circ_m$) above inputs in
the condition of a schema $f''$ means that \emph{any subset of
  inputs thus marked can `switch places' without affecting the
  automaton's transition.}
The index of the position-free symbol, when necessary, is used to
differentiate among distinct subsets of `permutable' inputs.
A two-symbol schema $f''_{\theta}$ redescribes a set $\Theta_{\theta} \equiv
\{f_{\alpha}: f_\alpha \rightarrowtail f''_{\theta}\}$ of LUT entries
of $x$;
it also redescribes a
subset $\Theta'_\theta \subseteq F'$  of wildcard schemata.

A two-symbol schema $f''_{\theta}$ captures \emph{permutation
  redundancy} in a set of wildcard schemata $\Theta'_\theta$.
More specifically, it identifies subsets of input states whose
permutations do not affect the truth value of the condition, leaving
the automaton's transition unchanged.
In group theory, a permutation is defined as a bijective mapping of a
non-empty set onto itself; a permutation group is any set of permutations of a set.
Permutation groups that consist of \emph{all} possible permutations of
a set are known as \emph{symmetric groups} under permutation
\cite{Wallace:1998ks}.
For Boolean functions in general, the study of
permutation/symmetric groups dates back to Shannon
\cite{Shannon:1938fk} and McCluskey \cite{MCCLUSKEY:1956nx} (see also
\cite{Kravets:2000fk}).

Two-symbol schemata identify subsets of wildcard schemata that form
symmetric groups.
We refer to each such subset of input states that can permute in a
two-symbol schema -- those marked with the same position-free symbol
-- as a \emph{group-invariant enput}.
Note that a group-invariant enput may include wildcard symbols marked
with a position-free symbol.
More formally, a two-symbol schema $f''$ can be expressed as a logical
conjunction of enputs -- literal or group-invariant.
Let us denote the set of literal enputs on the condition of $f''$ by $X_\ell \subseteq X$ -- the non-wildcard
inputs not marked with the position-free symbol. For simplicity,
$n_\ell = \left|X_\ell\right|$.

A group-invariant enput $g$ is defined by (1) a subset of input
variables $X_g \subseteq X$ that are marked with an identical
position-free symbol, and (2) a \emph{permutation constraint} (a
bijective mapping) on $X_g$ defined by the expression $n_g = n^0_g +
n^1_g + n^{\#}_g$, where $n_g = \left|X_g\right|$, $n^0_g$ is the
number of inputs in $X_g$ in state $0$ (off), and $n^1_g$ is the
number of inputs in $X_g$ in state $1$ (on).
We further require that at least two of the quantities $n_g^0, n_g^1$
and $n_g^\#$ are positive for any group-invariant enput $g$.
We can think of these two required positive quantities as
\emph{subconstraints}; in particular, we define a group-invariant
enput by the two subconstraints $n_g^0, n_g^1$, since $n_g^\#$ is
always derivable from those two given the expression for the overall
permutation constraint.
This precludes the trivial case of subsets of inputs in the same state from
being considered a valid group-invariant enput -- even though they can
permute leaving the transition unchanged.
A two-symbol schema $f''$ has $n_\ell$ literal enputs and $\eta$
group-invariant enputs; each of the latter type of enputs is defined
by a distinct permutation constraint for $g = 1,..., \eta$.
An input variable whose truth value is the wildcard symbol in a given schema is never a literal enput (it is not essential by definition).
However, it can be part of a group-invariant enput, if it is marked with a position-free symbol.
Further details concerning the computation of wildcard and two-symbol
schemata are available in \emph{Supporting text S2}.

In our working example, the resulting two-symbol schema (see Figure
\ref{fig:schema_example}C) contains $n_\ell=2$ literal inputs: $X_\ell
= \{i_2=0, i_3=1\}$.
It also contains one ($\eta=1$) group-invariant enput $X_g = \{i_1,i_4,i_5,i_6\}
$ with size $n_g=4$ and subconstraints $n^0_g = 1 \wedge n^1_g =
1$.
This redescription reveals that the automaton's transition to \emph{on} is
determined only by a subset of its six inputs:
\emph{as long as inputs 2 and 3 are \emph{off} and \emph{on},
  respectively, and among the others at least one is \emph{on} and
  another is \emph{off}, the automaton will transition to \emph{on}}.
These minimal control constraints are not obvious in the original LUT
and are visible only after redescription.

We equate \emph{canalization} with redundancy. The more redundancy
exists in the LUT of automaton $x$, as input-irrelevance or
input-symmetry (group-invariance), the more canalizing it is, and
the more compact its two-symbol redescription is, thus $|F''| < |F|$.
In other words -- after redescription -- input and input-symmetry
redundancy in $F$ is removed in the form of the two symbols.
The input states that remain are essential to determine the
automaton's transition.
Below we quantify these two types of redundancy, leading to two new
measures of canalization. Towards that, we must first clearly separate
the two forms of redundancy that exist in 2-symbol schemata.
The \emph{condition} of a two-symbol schema $f''$ with a single
group-invariant enput $g$ -- such as the one in Figure
\ref{fig:schema_example}C -- can be expressed as:

\begin{equation}
\displaystyle{
\bigwedge_{i_j \in X_\ell^0} \neg i_j \bigwedge_{i_j \in X_\ell^1}  i_j \wedge
\left(\sum_{i_j \in X_g} \neg i_j \geq n^0_g \right) \wedge \left(\sum_{i_j \in
    X_g} i_j \geq n^1_g \right)}
\label{si:canalogic}
\end{equation}

\noindent where $X_\ell^0$ is
the set of literal enputs that must be
\emph{off}, and $X_\ell^1$ is the set of literal enputs that must be
\emph{on} (thus $X_\ell = X_\ell^1 \cup X_\ell^0)$.
This expression separates the contributions (as conjunctions) of the
literal enputs, and each subconstraint of a group-invariant enput.
Since we found no automaton in the target model (see below) with
schemata containing more than one group-invariant enput, for simplicity and
without lack of generality, we present here only this case
($\eta=1$).
See \emph{Supporting text S3} for the general expression that accounts for
multiple group-invariant enputs ($\eta > 1$).

All possible transitions of $x$
to \emph{on} are described by a set $F_1''$ of two-symbol schemata.
This set can also be expressed in a DNF, where each disjunction
clause is given by Expression \ref{si:canalogic} for all schemata $f''
\in F_1''.$
Transitions to \emph{off} are defined by the negation of such DNF
expression: $F_0'' \equiv \left\{\neg f'',  \forall f'' \in F_1'' \right\}$.
Canalization of an automaton $x$ is now characterized in
terms of two-symbol schemata that capture two forms of redundancy: (1) \emph{input-irrelevance} and (2)
\emph{input-symmetry} (group-invariance).
We next describe the procedure to compute 2-symbol schemata for a an
automaton $x$.
Readers not interested in the algorithmic details of
this computation can safely move to the next subsection.

The procedure starts with the set of wildcard schemata $F'$ obtained
via the first step of the Quine \& McCluskey algorithm
\cite{Quine:1955ec} (see above).
The set $F'$ is then partitioned into subsets $H'_i$ such that,

\[
F' = \bigcup_i H'_i
\]

\noindent where each $H'_i$ contains every schema $x' \in F'$ with
equal number of zeroes ($n^0$), ones ($n^1$), and wildcards ($n^\#$),
with $n^0 + n^1 + n^\# = k$. In other words, the $H'_i$ are
equivalence classes induced on $F'$ by $n^0$, $n^1$, and $n^\#$.
This is a necessary condition for a set of wildcard schemata to form a
symmetric group.
The algorithm then iterates on each $H'_i$, checking if it contains a
symmetric group; i.e. if it contains wildcard schemata with all the
permutations of the largest set of inputs variables possible.
If it does, it marks those input variables as a group-invariant enput
in $H'_i$ and moves to another subset $H'_j$.
If it does not, then it checks for symmetric groups in smaller sets of
input variables within each set $H'_i$. It does this by iteratively
expanding the search space to include all subsets of $H'_i$ with
cardinality $|H'_i|-1$.
The procedure is repeated, if no symmetric groups are found, until the
subsets contain only one wildcard schema.

Although several heuristics are implemented to prune the search space,
the algorithm is often not suitable for exhaustively searching
symmetric groups in large sets of schemata.
However, the individual automata found in models of biochemical regulation and
signalling networks typically have a fairly low number of inputs.
Therefore, schema redescription of their LUT leads to manageable sets
of wildcard schemata, which can be exhaustively searched for symmetric
groups. Indeed, as shown below, all automata in the SPN model have
been exhaustively redescribed into two-symbol schemata.
For additional details on the computation of schemata see
\emph{Supporting text S2}.

\subsection*{Quantifying canalization: effective connectivity and input symmetry}
%\addcontentsline{toc}{subsection}{Quantifying canalization}

Schemata uncover the `control logic' of automata by making the
smallest input combinations that are necessary and sufficient to
determine transitions explicit.
We  equate canalization with the redundancy present in this
control logic: the smaller is the set of inputs needed to control an
automaton, the more redundancy exists in its LUT and the more
canalizing it is.
This first type of canalization is quantified by computing the mean
number of unnecessary inputs of automaton $x$, which we refer to as
\emph{input redundancy}.
An upper bound is given by,

\begin{equation}
\overline{k}_{\textrm{r}}(x) = \frac{  \displaystyle \sum_{f_{\alpha} \in F}
 \max_{ {\theta : f_\alpha \in \Theta_\theta}}\left(n_\theta^\# \right)}{|F|}
\label{upper_k_red}
\end{equation}

\noindent and a lower bound is given by:

\begin{equation}
\underline{k}_{\textrm{r}}(x) = \frac{  \displaystyle \sum_{f_{\alpha} \in F}
 \min_{ {\theta : f_\alpha \in \Theta_\theta}} \left(n_\theta^\# \right)}{|F|}
\label{lower_k_red}
\end{equation}

These expressions compute a mean number of
irrelevant inputs associated with the entries of the LUT $F$. The
number of irrelevant inputs in a schema $f''_{\theta}$ is the number
of its wildcards $n_{\theta}^{\#}$.
Because each entry $f_{\alpha}$ of $F$ is
redescribed by one or more schemata $f''_{\theta}$, there are various
ways to compute a characteristic number of irrelevant inputs
associated with the entry, which is nonetheless bounded by the maximum
and minimum number of wildcards in the set of schemata that redescribe $f_{\alpha}$.
Therefore, the expressions above identify all schemata $f''_{\theta}$
whose set of redescribed entries $\Theta_{\theta}$ includes
$f_{\alpha}$.
The upper (lower) bound of input redundancy, Equation
\ref{upper_k_red} (Equation \ref{lower_k_red}), corresponds to
considering the maximum (minimum) number of irrelevant inputs found
for all schemata $f''_{\theta}$ that redescribe entry $f_{\alpha}$ of
the LUT -- an optimist (pessimist) quantification of this type of
canalization.
Notice that input redundancy is not an estimated
  value. Also, it weights equally each entry of the LUT, which is the
  same as assuming that every automaton input is equally likely.

Here we use solely the upper bound, which we refer to
henceforth simply as \emph{input redundancy} with the notation
$k_{\textrm{r}}(x)$.
This means that we assume that the most redundant schemata are always
accessible for control of the automaton.
We will explore elsewhere the range between the bounds, especially in
regards to predicting the dynamical behaviour of BNs.
The range for input redundancy is $ 0 \le
k_{\textrm{r}}(x) \le k$, where $k$ is the number of inputs of $x$.
When $k_{\textrm{r}}(x) = k$ we have full input irrelevance, or
maximum canalization, which occurs only in the case of frozen-state
automata.
If $k_{\textrm{r}}(x) = 0$, the state of every
input is always needed to determine the transition and we have no
canalization in terms of input redundancy.

In the context of a BN, if some inputs of a node $x$ are irrelevant from a control logic perspective,
then its \emph{effective} set of inputs is smaller than its in-degree
$k$.
We can thus infer more about effective control in a BN than what
is apparent from looking at structure alone (see analysis of
macro-level control below).
A very intuitive way to quantify such effective control, is by
computing the mean number of inputs needed to determine the
transitions of $x$, which we refer to as its
\emph{effective connectivity}:

\begin{equation}
k_{\textrm{e}}(x) = k(x) - k_{\textrm{r}}(x)
\label{lower_eff_k}
\end{equation}

\noindent whose range is $ 0 \le k_{\textrm{e}}(x) \le k$. In this
case, $k_{\textrm{e}}(x) = 0$ means full input irrelevance, or maximum
canalization, and $k_{\textrm{r}}(x) = k$, means no canalization.

The type of canalization quantified by the input redundancy and
effective connectivity measures does not include the form of
permutation redundancy entailed by group-invariant enputs.
Yet this is a genuine form of redundancy involved in canalization, as
in the case of nested canalization \cite{Kauffman:2003ys}, since it
corresponds to the case in which different inputs can be
\emph{alternatively} canalizing.
The two-symbol schema redescription allows us to measure this form of
redundancy by computing the mean number of inputs that participate in
group-invariant enputs, easily tallied by the occurrence of the
position-free symbol ($\circ$) in schemata.
Thus we define a measure of \emph{input symmetry} for an automaton
$x$, whose upper-bound is given by

\begin{equation}
\overline{k}_{\textrm{s}}(x) = \frac{  \displaystyle \sum_{f_{\alpha} \in F}
 \max_{ {\theta : f_\alpha \in \Theta_\theta}}\left( n_\theta^\circ \right) }{|F|}
\label{eff_sym_upper}
\end{equation}

\noindent and a lower-bound by,

\begin{equation}
\underline{k}_{\textrm{s}}(x) = \frac{  \displaystyle \sum_{f_{\alpha} \in F}
 \min_{ {\theta : f_\alpha \in \Theta_\theta}}\left( n_\theta^\circ \right) }{|F|}
\label{eff_sym_lower}
\end{equation}

\noindent where $n_\theta^\circ$ is the number of position-free symbols in schema $f''_{\theta}$.

The upper bound of input symmetry (Equation \ref{eff_sym_upper})
corresponds to considering an optimist quantification of this type of
canalization.
Here we use solely the upper bound, which we refer to
henceforth simply as input symmetry and denote by ${k}_{\textrm{s}}(x)$.
Again, the assumption is that the most redundant schemata are always accessible for control of the automaton.
The range for input symmetry is $0 \le {k}_{\textrm{s}}(x) \le k$.
High (low) values mean that permutations of input states are likely
(unlikely) to leave the transition unchanged.

Canalization in automata LUTs -- the micro-level of networks of
automata -- is then quantified by two types of redundancy: \emph{input
  redundancy} using $k_{\textrm{r}}(x)$ and \emph{input symmetry} with
$k_{\textrm{s}}(x)$.
%
%In Figure \ref{fig:cm_example}C, the canalization of running example
%automaton (from Figure \ref{fig:schema_example}) is shown and discussed.
%
To be able to compare the canalization in automata with distinct
numbers of inputs, we can compute \emph{relative} measures of
canalization:

\begin{equation}
k_{\textrm{r}}^{*}(x) = \frac{k_{\textrm{r}}(x)}{k(x)}; \quad  k_{\textrm{s}}^{*}(x) = \frac{k_{\textrm{s}}(x)}{k(x)}
\label{relative_canalization_measures}
\end{equation}

\noindent the range of which is $[0, 1].$
Automata transition functions can have different amounts of each form
of canalization, which allows us to consider four broad canalization
classes for automata:
\emph{class A} with high $k_{\textrm{r}}(x)$ and high
$k_{\textrm{s}}(x)$, \emph{class B} with high $k_{\textrm{r}}(x)$ and
low $k_{\textrm{s}}(x)$, \emph{class C} with low $k_{\textrm{r}}(x)$
and high $k_{\textrm{s}}(x)$, and \emph{class D} with low
$k_{\textrm{r}}(x)$ and low $k_{\textrm{s}}(x)$.
We will explore these classes in more detail elsewhere.
Below, these measures are used to analyse micro-level canalization in the
SPN model and discuss the type of functions encountered.
Before that, let us introduce an alternative representation of the
canalized control logic of automata, which allows us to compute
network dynamics directly from the parsimonious information provided
by schemata.

\subsection*{Network representation of a schema}
%\addcontentsline{toc}{subsection}{Network representation of a schema}

Canalization in an automaton, captured by a set of schemata, can also
be conveniently represented as a McCulloch \& Pitts threshold network
-- introduced in the 1940s to study computation in
interconnected simple logical units \cite{McCulloch:1943uq}.
These networks consist of binary units that can transition from
quiescent to firing upon reaching an activity threshold
$(\tau)$ of the firing of input units.
To use this type of network to represent two-symbol schemata we resort
to two types of units.
One is the \emph{state unit} (s-unit), which represents an input
variable in a specific Boolean state; the other is the \emph{threshold unit}
(t-unit) that implements the condition that causes the
automaton to transition.
Two s-units are always used to represent the (Boolean) states of any
input variable that participates as enput in the condition of an
automaton $x$: one fires when the variable is \emph{on} and the other
when it is \emph{off}.
To avoid contradiction, the two s-units for a given variable cannot
fire simultaneously.
Directed fibres link (source) units to (end) units, propagating a
pulse -- when the source unit is firing -- that contributes to the firing of
the end unit.
The simultaneous firing of at least $\tau$ (threshold)
incoming s-units into a t-unit, causes the latter to fire.

In the example automaton in Figure \ref{fig:schema_example}, the set
of schemata $F''$ contains only one schema. This schema can be directly
converted to a (2-layer) McCulloch \& Pitts network.
This conversion, which is possible due to the separation of
subconstraints given by Expression \eqref{si:canalogic}, is shown in Figure
\ref{fig:CM_translation_straight} and explained in its caption.
Note that in the McCulloch \& Pitts representation, the transition of
the automaton is determined in two steps.
First, a \emph{layer} of threshold units is used to check that the
literal and group-invariant constraints are satisfied; then, a second
layer -- containing just one threshold unit -- fires when every
subconstraint in Expression \eqref{si:canalogic} has been simultaneously
satisfied, determining the transition.
This means that in this network representation each schema with
literal enputs and at least a group-invariant enput requires two
layers and three t-units.
Since in McCulloch \& Pitts networks each threshold unit
has a standard delay of one time step, this network representation of
a schema takes two time steps to compute its transition.
We introduce an alternative threshold network representation of a
two-symbol schema $f''$ that only requires a single t-unit and
takes a single time delay to compute a transition.
We refer to this variant as the \emph{Canalizing Map} of a schema or
CM for short.
A CM is essentially the same as a McCulloch and Pitts network,
with the following provisos concerning the ways in which s-units and
t-units can be connected:

\begin{enumerate}
\item{Only one fibre originates from each s-unit that can participate
    as enput in $f''$ and it must always end in the t-unit used to
    encode $f''$.}
\item{The fibre that goes from a s-unit to the t-unit can \emph{branch
      out} into several fibre endings.
    This means that if the s-unit is firing, a pulse propagates
    through its outgoing fibre and through its branches. Branching
    fibres are used to capture group-invariant enputs, as we explain
    later.}
\item{Branches from distinct s-units can \emph{fuse} together
    into a single fibre ending -- the fused fibre increases the end
    t-unit's firing activity by one if at least one of the fused
    fibres has a pulse.}
\item{A fibre that originates in a t-unit encoding a schema $f''$ must
    end in a s-unit that corresponds to the automaton
    transition defined by $f''$.}
\end{enumerate}

\begin{figure}
\includegraphics{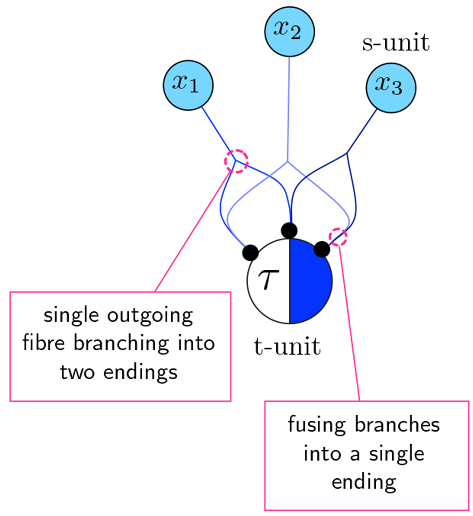}
\caption{{\bf Elements of a Canalizing Map}.  Every s-unit is a
  circle, labelled according the automaton's input it represents and
  coloured according to its state: black is \emph{on} and white is
  \emph{off} (here we use light-blue for a generic state). The
  t-unit (schema) is represented using a larger circle.  One of its
  halves is coloured, and the other labelled with the t-unit's
  threshold $\tau$. Fibres can be single, or branched. In this example
  there are branching fibres only, where fibre fusions represent all
  possible combinations of two out of the three s-units.}
\label{fig:cm_elements}
\end{figure}

Figure \ref{fig:cm_elements} depicts the elements of a single schema's
CM.  Table \ref{tab:TN_Rules_Wiring} summarizes the rules that apply
to the interconnections between units.
Transitions in CMs occur in the same way as in standard McCulloch \&
Pitts networks.
Once sufficient conditions for a transition are observed at some time
$t$, the transition occurs at $t+1$.
The firing (or not) of t-units is thus assumed to have a standard
delay (one time-step), identical for all t-units.
Note that in CMs, s-units can be regarded as a special type of t-unit with
threshold $\tau=1$ that send a pulse through their outgoing fibres
instantaneously.
Next we describe the algorithm to obtain the CM representation of a
schema. Readers not interested in the algorithmic details
of this computation can safely bypass the next subsection.

\subsubsection*{Algorithm to obtain the canalizing map of a schema}
%\addcontentsline{toc}{subsubsection}{Algorithm to obtain the canalizing map of a schema}

\begin{figure}
\includegraphics{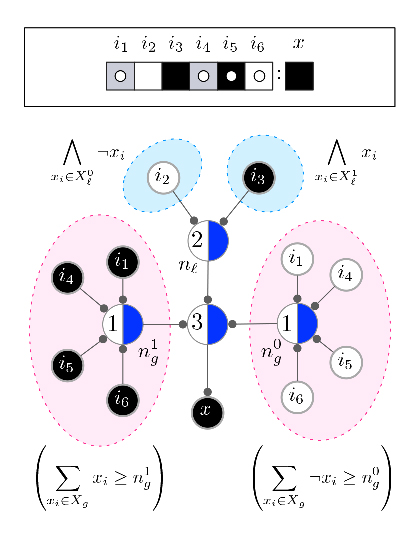}
\caption{{\bf McCulloch \& Pitts representation of Expression
    \eqref{si:canalogic}.}  The conjunction clauses in Expression
  \eqref{si:canalogic} for the example automaton $x$ are directly
  mapped onto a standard McCulloch \& Pitts network with two
  layers. On one layer the two literal enputs are accounted for by a
  threshold unit (at the top) with threshold $\tau=n_\ell = 2$. There
  is also a group-invariant enput with permutation subconstraints on
  both Boolean states. Two threshold units on the same layer are used
  to capture these. The threshold unit on the left accounts for the
  permutation subconstraint $n_g^1 = 1$. It thus has as incoming
  s-units the inputs $x_i \in X_g : x_i =1$ and threshold $\tau =
  n_g^1 = 1$. In a similar way, the threshold unit on the right
  accounts for the subconstraint $n_g^0=1$. When all the constraints
  (literal and group-invariant) are satisfied, the last threshold unit
  (second layer) `fires' causing the transition to \emph{on}.. }
\label{fig:CM_translation_straight}
\end{figure}

\begin{figure}
\includegraphics{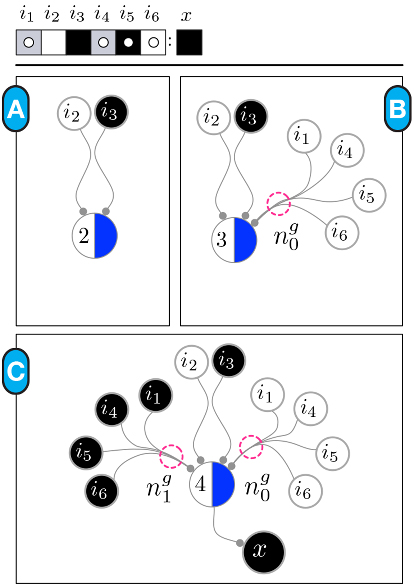}
\caption{{\bf Canalizing map of example automaton $x$ characterized by
    a single schema.} (A) Since $f''$ (shown on top) has $n_\ell=2$,
  the corresponding s-units for literal enputs $x_i \in X_\ell$ are
  directly linked to the t-unit for $f''$ with single fibres; $\tau =
  n_\ell = 2$. (B) Adding the subconstraint $n_g^0=1$ of the
  group-invariant enput $X_g = \{i_1,i_4,i_5,i_6\}$. In this case,
  $n_g - (n_g^0 -1) = n_g = 4$, so there is only one subset $S_i
  \subseteq S$ and thus a single branch from each s-unit in the
  group-invariant, fused into a single ending. The threshold becomes
  $\tau = n_\ell + {n_g \choose n_g^0 -1} = 2 + {4 \choose 0} =
  3$. (C) Finally, we add the second subconstraint $n_g^1 =1$ of the
  group-invariant enput $X_g$, which has the same properties as the
  subconstraint integrated in (B). The final threshold of the t-unit
  is given by \eqref{eq:tau}, therefore $\tau = n_\ell + {n_g \choose
    n_g^0 -1} + {n_g \choose n_g^1 -1} = 2 + {4 \choose 0} + {4
    \choose 0}= 4$. Notice that only the input-combinations that
  satisfy the constraints of Expression \eqref{si:canalogic} for $f''$
  can lead to the firing of the t-unit; in other words, the canalizing
  map is equivalent to schema $f''$.}
\label{fig:final_method_example_x}
\end{figure}

Given a 2-symbol schema $f''$,
there are two steps involved in producing its CM representation.
The first is connecting s-units to the t-unit for $f''$ in such a way
that it fires, if and only if, the constraints of $f''$ -- defined by
Expression \eqref{si:canalogic} -- are satisfied.
The second step is determining the appropriate firing threshold $\tau$
for the t-unit.
If the schema does not have group-invariant enputs, the conversion is
direct as for the standard McCulloch \& Pitts network -- see Figure
\ref{fig:CM_translation_straight}:
The s-units corresponding to literal enputs $i_j \in X_\ell$ are
linked to the t-unit using a single fibre from each s-unit to the
t-unit, which has a threshold $\tau = n_\ell$.
If the schema has a group-invariant enput, its subconstraints are
implemented by branching and fusing fibres connecting the s-units
and the t-unit.
In cases such as our example automaton $x$ (Figures
\ref{fig:schema_example} and \ref{fig:CM_translation_straight}) where
the subconstraints $n_g^0 = n_g^1 = 1$, the solution is still simple.
To account for subconstraint $n_g^0$, it is sufficient to take an
outgoing fibre from each of the s-units $i_j \in X_g : i_j=0$ and fuse them into a single fibre ending.
Therefore, if at least one of these s-units is firing, the fused fibre
ending transmits a single pulse to the t-unit, signalling that the
subconstraint has been satisfied.
Increasing the t-unit's threshold by one makes the t-unit respond to
this signal appropriately.
The same applies for subconstraint $n_g^1$, using a similar wiring for s-units $i_j \in X_g : i_j=1$.
The final threshold for the t-unit that captures the example schema of Figure \ref{fig:schema_example} is
thus $\tau = n_\ell + n_g^0 + n_g^1 = 2 + 1 + 1 = 4$, as shown in Figure
\ref{fig:final_method_example_x}C.

The case of general group-invariant constraints is more
intricate.
Every literal enput $i_j \in X_\ell$ is linked to the t-unit via a
single fibre exactly as above.
Afterwards, the subconstraints $n_g^0$ and $n_g^1$ of a
group-invariant enput $g$ are treated separately and consecutively.
Note that for every input variable $i_j$ in the set $X_g$ of symmetric input variables, there are two s-units: one representing $i_j$ in state $0$ and another in state $1$.
To account for subconstraint $n_g^0$ on the variables of set $X_g$, let $S \subseteq X_g$ be the
set of s-units that represent the variables of the
group-invariant enput that can be in state $0$, where $|X_g| = n_g$.
%
%: $i_j \in X_g : i_j =0$.
%
Next, we identify all possible subsets of $S$, whose cardinality is
$n_g -(n_g^0-1)$. That is, $\bm{S} = \left\{ S_i : S_i
\subset S \wedge |S_i| = n_g -(n_g^0-1)\right\}$.
For each subset $S_i \in \bm{S}$, we take an outgoing fibre
from every s-unit in it and fuse them into a single fibre
ending as input to the schema t-unit.
After subconstraint $n_g^0$ is integrated this way, the threshold of the
t-unit is increased by,

\begin{equation}
|\bm{S}| = {n_g \choose n_g - (n_g^0 -1)}  = {n_g \choose n_g^0 -1}
\end{equation}

\noindent This procedure is repeated for the subconstraint $n_g^1$ on $X_g$.
The final threshold of the t-unit is,

\begin{equation}
\tau = n_\ell + {n_g \choose n_g^0 -1} + {n_g \choose n_g^1 -1}
\label{eq:tau}
\end{equation}

\begin{figure}
\includegraphics{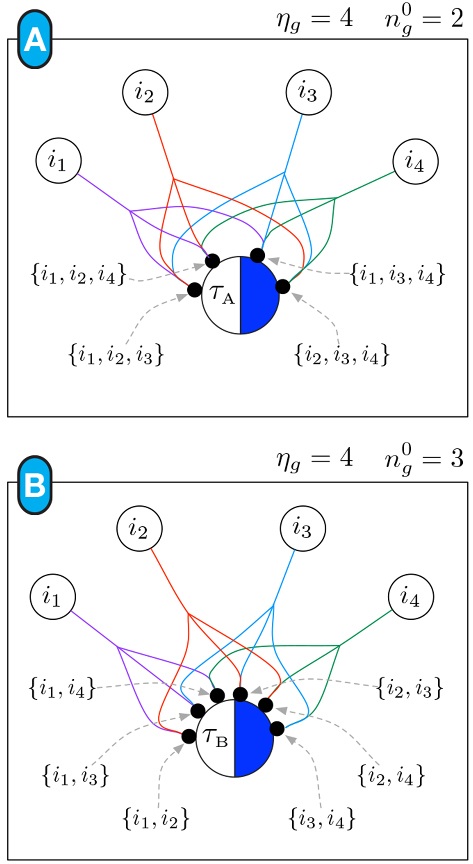}
\caption{{\bf Procedure for obtaining the canalizing map of a
    group-invariant subconstraint.} (A) subconstraint defined by
  $n_g^0=2$, where $n_g=4$. The first step is to consider the s-units
  (in state 0) for the four input variables in the group invariant
  enput $X_g = \{i_1,i_2,i_3,i_4\}$. Next we identify all the subsets
  $S_i$ of these s-units containing $n_g - (n_g^0 -1) = 3$ s-units:
  $\{i_1,i_2,i_3\}, \{i_1,i_2,i_4\},\{i_1,i_3,i_4\}, \{i_2,i_3,i_4\}$
  (shown with dotted arrows). From every s-unit in each such subset
  $S_i$, we take an outgoing fibre to be joined together into a single
  fibre ending as input to the t-unit. Finally, we increase the
  threshold of the t-unit by the total number of subsets, that is
  $\tau_A = {n_g \choose n_g^0 -1} = {4 \choose 4} = 4$. (B) An
  example of the same procedure but for $n_g^0=3$ and $n_g=4$: $\tau_B
  = {n_g \choose n_g^0 -1} = {4 \choose 2} = 6$.}
\label{fig:final_method_example23}
\end{figure}

This algorithm is illustrated for the integration of two example
subconstraints in Figure \ref{fig:final_method_example23}; in Figure
\ref{fig:final_method_example_x}, the case of the only schema
describing the transitions to \emph{on} of running example automaton
$x$ is shown.
Further details concerning this procedure are provided in 
\emph{Supporting text S3}.

\subsubsection*{The canalizing map of an automaton}
%\addcontentsline{toc}{subsubsection}{The canalizing map of an automaton}

The algorithm to convert a single schema $f''$ to a CM is subsequently used to
produce the CM of an entire Boolean automaton $x$ as follows:
Each schema $f'' \in F''$ is converted to its CM representation.
Each state of an input variable is represented by a single s-unit in the
resulting threshold network.
In other words, there is a maximum of two
s-units (one for state $0$ and one for state $1$) for each input
variable that is either a literal enput or participates in a
group-invariant enput of $x$.
The resulting threshold network is the canalizing map of $x$.
The connectivity rules of automata CMs include the
following provisos:

\begin{enumerate}
\item{Every s-unit can be connected to a single t-unit with a single
    outgoing fibre, which can be single or have branches. }

\item{Therefore, the number of outgoing fibres coming out of a s-unit
    (before any branching) corresponds to the number of schemata $f''
    \in F''$ in which the respective variable-state participates as an
    enput. If such a variable is included in a group-invariant enput,
    then the fibre may have branches.}

\item{Any subset set of t-units with threshold $\tau=1$ for the same
    automaton transition ($x=0$ or $x=1$) are merged into a single
    t-unit (also with $\tau=1$), which receives all incoming fibres of
    the original t-units. In such scenario, any fused branches can
    also be de-fused into single fibres. Note that this situation
    corresponds to schemata that exhibit nested canalization, where
    one of several inputs settles the transition, but which do not
    form a symmetric group.}

\end{enumerate}

\begin{figure}
\includegraphics{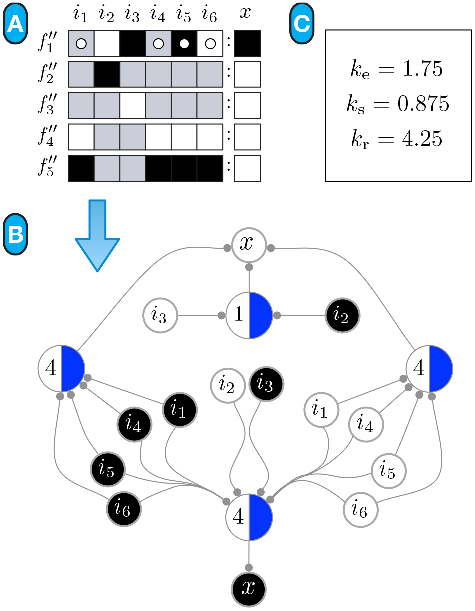}
\caption{{\bf Canalizing Map of automaton $x$}. (A) complete set of
  schemata $F''$ for $x$, including the transitions to
  \emph{on} shown in Figure \ref{fig:schema_example} and the
  transitions of \emph{off} (the negation of the first).(B) canalizing
  map; t-units for schemata $f''_2$ and $f''_3$ were merged into a
  single t-unit with threshold $\tau=1$ (see main text). (C) effective
  connectivity, input symmetry and input redundancy of $x$.}
\label{fig:cm_example}
\end{figure}

The CM of $x$ can be constructed from the subset of schemata $F_1''$
(the conditions to \emph{on}), or $F_0''$ (the conditions to
\emph{off}).
When the conditions are not met for convergence to \emph{on}, one is
guaranteed convergence to \emph{off} (and vice-versa).
However, since we are interested in exploring scenarios with
incomplete information about the states of variables in networks of
automata rather than a single automaton (see below), we construct the
CM of a Boolean automaton $x$ including all conditions, that is using
$F'' \equiv F_1'' \cup F_0''$. This facilitates the analysis of
transition dynamics where automata in a network can transition to
either state.
Figure \ref{fig:cm_example} depicts the complete CM of the
example automaton $x$ described in Figure \ref{fig:schema_example} --
now including also its transitions to \emph{off}.

By uncovering the enputs of an
automaton, we gain the ability to compute its transition with
\emph{incomplete information} about the state of every one of its
inputs.
For instance, the possible transitions of the automaton in Figure
\ref{fig:schema_example} are fully described by the CM (and schemata)
in Figure \ref{fig:cm_example}; as shown, transitions can be
determined from a significantly small subset of the input variables in
specific state combinations.
For instance, it is sufficient to observe $i_3=0$ to know that
automaton $x$ transitions to \emph{off}.
If $x$ was used to model the interactions that lead a gene to be
expressed or not, it is easy to see that to down-regulate its
expression, it is sufficient to ensure that the regulator $i_3$ is not expressed.
This is the essence of canalization: the transition of an automaton is
controlled by a small subset of input states.
In the macro-level canalization section below, we use the CM's ability
to compute automata transitions with incomplete information to
construct an alternative portrait of network dynamics, which we use in
lieu of the original BN to study collective dynamics.
Let us first apply our micro-level methodology to the SPN model.

\subsection*{Micro-level canalization in the SPN model}
%\addcontentsline{toc}{subsection}{Micro-level canalization in the SPN}

\begin{figure}
\includegraphics{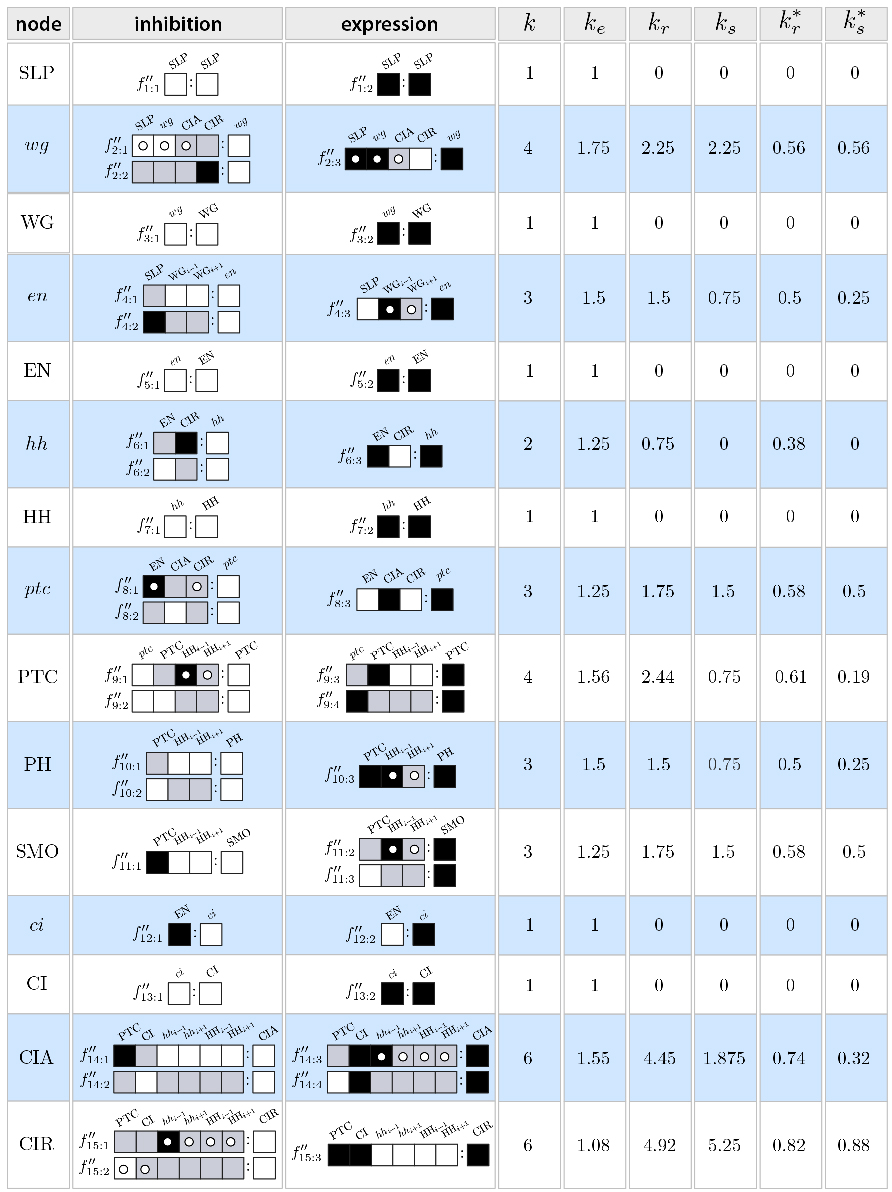}
\caption{{\bf Micro-level canalization for the Automata in the SPN
    model}.Schemata for inhibition (transitions to off) and expression
  (transitions to on) are shown for each node (genes or proteins) in
  model. In-degree ($k$), input redundancy ($k_{\textrm{r}}$),
  effective connectivity ($k_{\textrm{e}}$), and input symmetry
  ($k_{\textrm{s}}$) are also shown.}
\label{fig:spn_schemata}
\end{figure}

\begin{figure}
\includegraphics{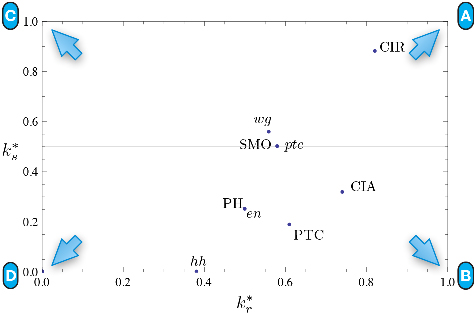}
\caption{{\bf Quantification of canalization in the SPN
    automata}. Relative input redundancy is measured in the horizontal
  axis ($k_{\textrm{r}}^{*}$) and relative input symmetry is measured
  in the vertical axis ($k_{\textrm{r}}^{*}$). Most automata in the
  SPN fall in the class II quadrant, showing that most canalization is
  of the input redundancy kind, though nodes such as CIR and $wg$
  display strong input symmetry too.}
\label{fig:spn_canalization_qualitative}
\end{figure}

\begin{figure}
\includegraphics[width=10.5cm]{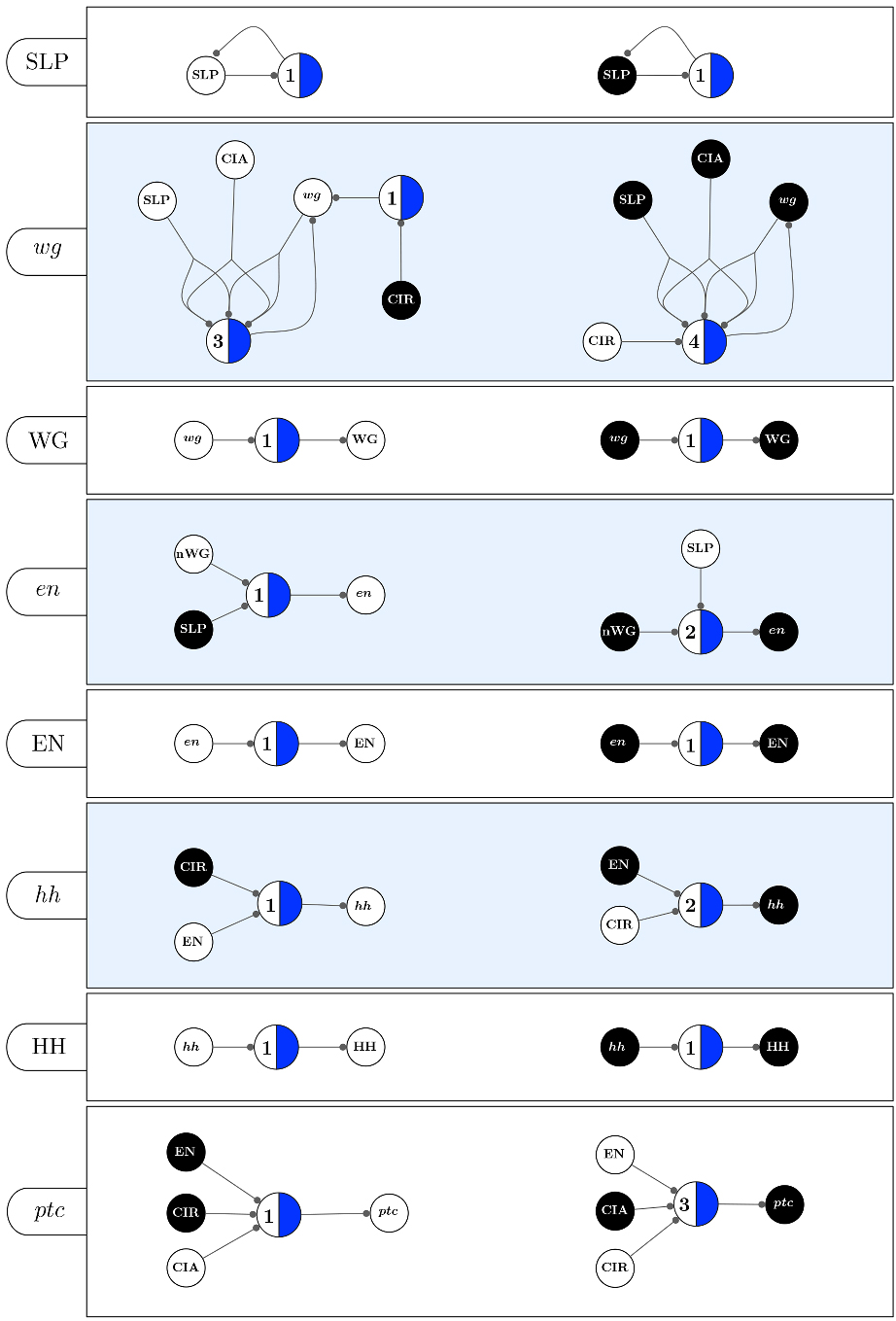}
\caption{{\bf Canalizing Maps of individual nodes in the SPN model
    (part 1).} The set of schemata for each automaton is converted
  into two CMs: one representing the minimal control logic for its
  inhibition, and another for its expression.
Note that $nX$ denotes the state of node $X$ in both neighbour cells:
$\neg nX \Leftrightarrow \neg X_{i-1} \wedge \neg X_{i+1}$ and $nX
\Leftrightarrow X_{i-1} \vee X_{i+1}$, where $X$ is one of the
spatial-signals $hh$, HH, or WG (see text).}
\label{fig:spn_cm_p1}
\end{figure}

\begin{figure}
\includegraphics[width=10.5cm]{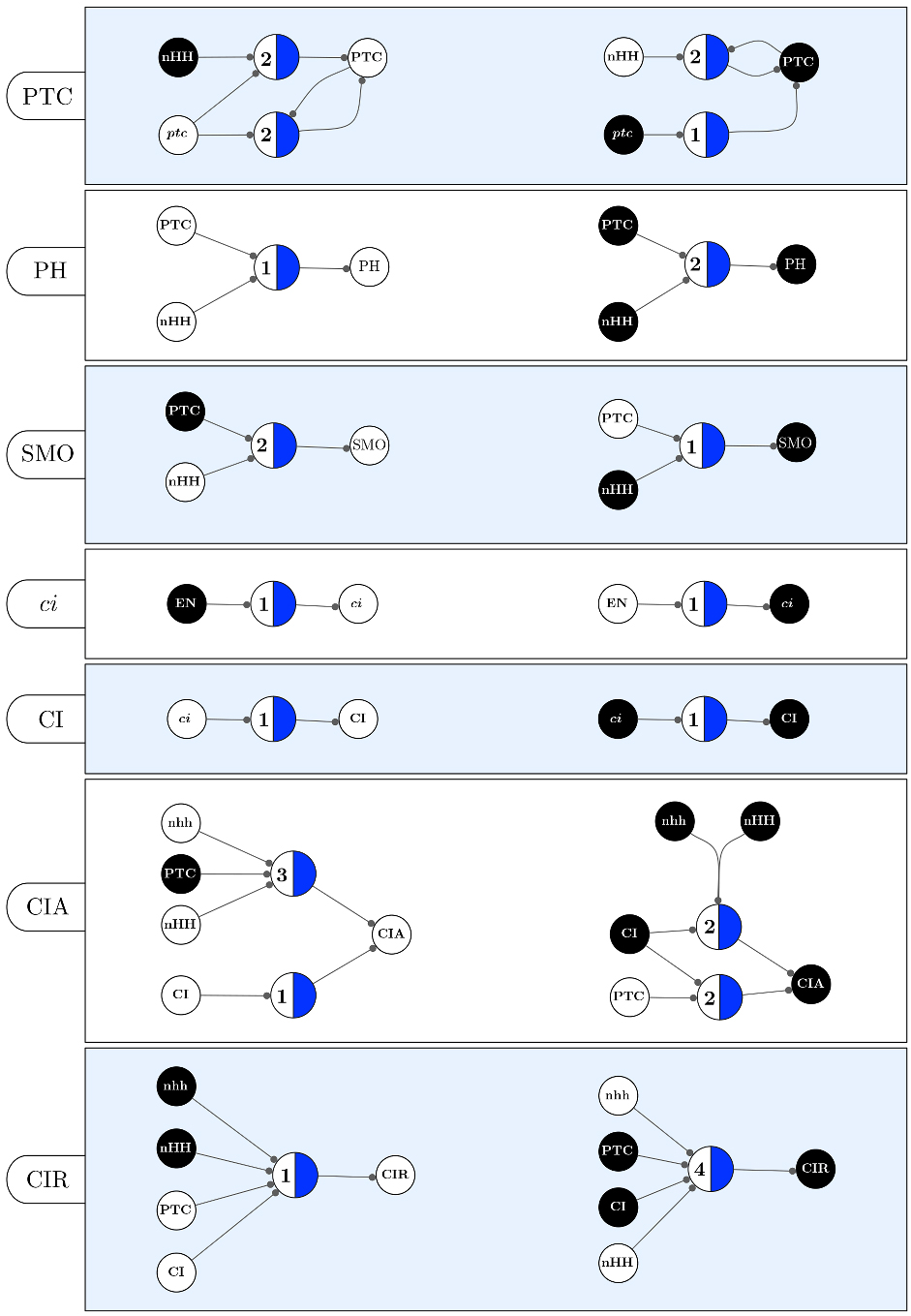}
\caption{{\bf Canalizing Maps of individual nodes in the SPN model
    (cont).} The set of schemata for each automaton is converted into two CMs: one representing the minimal control
logic for its inhibition, and another for its expression.
Note that $nX$ denotes the state of node $X$ in both neighbour cells:
$\neg nX \Leftrightarrow \neg X_{i-1} \wedge \neg X_{i+1}$ and $nX
\Leftrightarrow X_{i-1} \vee X_{i+1}$, where $X$ is one of the
spatial-signals $hh$, HH, or WG (see text).}
\label{fig:spn_cm_p2}
\end{figure}

The automata in the SPN fall in two categories: those that have a
single input ($k=1$), the analysis of which is trivial, namely, SLP,
WG, EN, HH, $ci$ and CI, and those with $k > 1$.
The two-symbol schemata and canalization measures for each automaton
in the SPN model are depicted in Figure \ref{fig:spn_schemata}; Figure
\ref{fig:spn_canalization_qualitative} maps the automata to their
canalization classes.
Schemata easily display all the sufficient combinations
of input states (enputs) to control the transitions of the
automata in this model, which represent the inhibition or expression
of genes and proteins.
Indeed, the resulting list of schemata allows analysts to quickly
infer how control operates in each node of the network.
Wildcard symbols (depicted in Figure \ref{fig:spn_schemata} as grey
boxes) denote redundant inputs.
Position-free symbols (depicted in Figure \ref{fig:spn_schemata} as
circles), denote `functionally equivalent' inputs; that is, sets of
inputs that can be alternatively used to ensure the same transition.
For example, for $wg$ to be expressed, SLP, the previous state of $wg$
(reinforcing feedback loop) and CIA can be said to be `functionally
equivalent', since any two of these three need to be expressed for
$wg$ to be expressed.
The several schemata that are listed for the expression or inhibition
of a specific node (genes and gene products), give experts alternative `recipes' available to
control the node according to the model -- and which can be
experimentally tested and validated.
Let us now present some relevant observations concerning micro-level
canalization in the SPN model:

\begin{enumerate}
\item{The inhibition of $wg$ can be attained in one of two ways: either two
    of the first three inputs (SLP, $wg$, CIA) are \emph{off}
    (unexpressed), or CIR is \emph{on} (expressed).
    The expression of $wg$ -- essential in the posterior cell of a
    parasegment to attain the wild-type expression pattern (Figure \ref{fig:spn_parasegment})-- is
    attained in just one way: CIR must be \emph{off} (unexpressed),
    and two of the other three inputs (SLP, $wg$, CIA) must be
    \emph{on} (expressed).
    Note the simplicity of this control logic vis a vis the $2^4 = 16$
    possible distinct ways to control $wg$ specified by its LUT, given
    that it is a function of 4 inputs.
    This control logic is also not obvious from the Boolean logic
    expression of node $wg$, as shown in Table
    \ref{tab:SPN_Logic_Table}; at the very least, the schemata
    obtained for $wg$ provide a more intuitive representation of
    control than the logical expression. Moreover, schema
    redescription, unlike the logical expression, allows us to directly
    quantify canalization.
    The control logic of this gene shows fairly high degree of both
    types of canalization: even though there are $k=4$ inputs, on
    average, only $k_e = 1.75$ inputs are needed to control the
    transition, and $k_s = 2.25$ inputs can permute without effect on
    the transition (see Figures \ref{fig:spn_schemata} and
    \ref{fig:spn_canalization_qualitative}); $wg$ is thus modelled by
    an automaton of class A.}
\item{The inhibition of CIR can be attained in one of two simple,
    highly canalized, ways: either one of its first two inputs (PTC,
    CI) is \emph{off} (unexpressed), or one of its four remaining
    inputs ($hh$ and $HH$ in neighbouring cells) is \emph{on}
    (expressed); all other inputs can be in any other state.
    The expression of CIR can be attained in only one specific,
    non-canalized, way: the first two inputs must be \emph{on}
    (expressed), and the remaining four inputs must be \emph{off}
    (unexpressed) -- a similar expression behaviour is found for $hh$
    and $ptc$.
    Note the simplicity of this control logic vis a vis the $2^6 = 64$
    possible distinct ways to control CIR specified by its LUT, given
    that it is a function of $6$ inputs.
    While, in this case, the control logic is also pretty clear from
    the original Boolean logic expression of node CIR (in Table
    \ref{tab:SPN_Logic_Table}), the schemata obtained for CIR provide
    a more intuitive representation of control and allows us to
    directly quantify canalization.
    CIR is a protein with a very high degree of both types of
    canalization: even though there are $k=6$ inputs, on average, only
    $k_e = 1.08$ inputs are needed to control the transition, and $k_s
    = 5.25$ inputs can permute without effect on the transition (see
    Figures \ref{fig:spn_schemata} and
    \ref{fig:spn_canalization_qualitative}).
    This high degree of both types of canalization, which is not
    quantifiable directly from the logical expression or the LUT, is
    notable in Figure \ref{fig:spn_canalization_qualitative}, where
    CIR emerges very clearly as an automaton of class A.}
\item{The control logic of CIA entails high canalization of the input
    redundancy kind. For instance, its inhibition can be achieved by a
    single one of its six inputs (CI \emph{off}) and its expression by
    two inputs only (PTC \emph{off} and CI \emph{on}). On the other
    hand, there is low canalization of the input symmetry kind,
    therefore CIA is modelled by an automaton in class B.}

\item{The expression of $en$ -- essential in the anterior cell of a
    parasegment to achieve the wild-type phenotype -- depends on the
    inhibition of (input node) SLP in the same cell, and on the
    expression of the wingless protein in at least one neighbouring
    cell.}

\item{Most automata in the model fall into canalization class B
    described above. CIR and $wg$ discussed above display greatest input
    symmetry, and fall in class A (see Figure
    \ref{fig:spn_canalization_qualitative})}.
\item{Looking at all the schemata obtained in Figure
    \ref{fig:spn_schemata}, we notice a consistent pattern for all
    spatial signals, $hh_{i \pm 1} , \textrm{HH}_{i \pm 1}$ and
    $\textrm{WG}_{i \pm 1}$. Whenever they are needed to control a
    transition (when they are enputs in the schemata of other nodes), either they are
    \emph{off} in both neighbouring cells, or they are \emph{on} in at least one of the neighbouring cells.
    For instance, for a given cell $i$, HH in neighbouring cells is
    only relevant if it is unexpressed in both cells ($\textrm{HH}_{i
      \pm 1} = 0$), or if it is expressed in at least one of them
    ($\textrm{HH}_{i - 1} = 1 \vee \textrm{HH}_{i + 1} = 1$).
    This means that the six nodes corresponding to spatial signals
    affecting a cell in a parasegment can be consolidated into just
    three \emph{neighbour nodes}, a similar consolidation of spatial
    signals was used previously by Willadsen \& Wiles
    \cite{Willadsen:2007hc} to simplify the spatial model into a
    single-cell non-spatial model.
In what follows, we refer to these spatial signals simply as $nhh$, nHH
and nWG.
If such a node is \emph{off} it means that the corresponding original
nodes are \emph{off} in both adjacent cells; if it is \emph{on} it
means that at least one of the corresponding original nodes in an
adjacent cell is \emph{on}.}
\item{Only PTC and $wg$ have feedback loops that are active after
    schema redescription, both for their inhibition
    and expression; these are self-reinforcing, but also depend on
    other enputs (see also Figures \ref{fig:spn_cm_p1} and \ref{fig:spn_cm_p2}).}
\end{enumerate}

Because this is a relatively simple model, some of the observations
about control, especially for nodes with fewer inputs, could be made
simply by looking at the original transition functions in Table
\ref{tab:SPN_Logic_Table}, since they are available as very simple
logical expressions -- this is the case of CIR, but certainly not $wg$
above.
However, the \emph{quantification} of canalization requires the
additional symbols used in schema redescription to identify
redundancy, which are not available in the original automata logical
expressions or their LUTs.
Moreover, the transition functions of automata in larger Boolean models
of genetic regulation and signalling are rarely available as simple
logical expressions, and nodes can be regulated by a large number of
other nodes, thus making such direct comprehension of control-logic difficult.
In contrast, since redescription uncovers canalization in the form of
input redundancy and symmetry, the more canalization exists, the more
redundancy is removed and the simpler will be the schemata
representation of the logic of an automaton.
This makes canalizing maps (CM) particularly useful, since they can be
used to visualize and compute the minimal control logic of automata.
The CMs that result from converting the schemata of each node in the
SPN to a threshold-network representation are shown in Figure
\ref{fig:spn_cm_p1} and Figure \ref{fig:spn_cm_p2}.
For a biochemical network of interest, such as the SPN or much larger
networks, domain experts (e.g. biomedical scientists and systems and
computational biologists) can easily ascertain the control logic of
each component of their model from the schemata or the corresponding
CMs.

In summary, there are several important benefits of schema
redescription of Boolean automata vis a vis the original Boolean logic
expression or the LUT of an automaton: (1) a parsimonious and
intuitive representation of the control logic of automata, since
\emph{redundancy is clearly identified} in the form of the two
additional symbols, which gives us (2) the ability to \emph{quantify}
all forms of canalization in the straightforward manner described
above; finally, as we elaborate next, the integration of the schema
redescription (or CMs) of individual automata in a network
(micro-level) allows us to (3) \emph{characterize macro-level
  dynamics} parsimoniously, uncovering minimal control patterns, robustness and
the modules responsible for collective computation in these networks.

\subsection*{Macro-level canalization and control in automata networks}
%\addcontentsline{toc}{subsection}{Macro-level canalization and control in automata networks}

After removing redundancy from individual automata LUTs in networks
(micro-level), it becomes possible to integrate their canalizing logic
to understand control and collective dynamics of automata networks
(macro-level).
In other words, it becomes feasible to understand how biochemical
networks process information collectively -- their emergent or
collective computation
\cite{Mitchell:2006fk,Peak:2004fk,Crutchfield:1995tm,Marques-Pita:2011uq,Rocha05}.

\subsubsection*{Dynamics canalization map and dynamical modularity}
%\addcontentsline{toc}{subsubsection}{DCM and modularity}

The CMs obtained for each automaton of a BN, such as the SPN model
(see Figures \ref{fig:spn_cm_p1} and \ref{fig:spn_cm_p2}), can be
integrated into a single threshold network that represents the control
logic of the entire BN.
This simple integration requires that (1) each automaton is
represented by two unique s-units, one for transition to \emph{on} and
another to \emph{off}, and (2) s-units are linked via t-units with
appropriate fibres, as specified by each individual CM. Therefore a
unique t-unit represents each schema obtained in the
redescription process.
This results in the \emph{Dynamics Canalization Map} (DCM) for the
entire BN.
Since the DCM integrates the CMs of its constituent automata, it can
be used to identify the \emph{minimal control conditions} that are sufficient
to produce transitions in the dynamics of the entire network.
Notice that when a node in the original BN undergoes a
state-transition, it means that at least one t-unit fires in the
DCM.
When a t-unit fires, according to the control logic of the DCM,
it can cause subsequent firing of other t-units.
This allows the identification of the \emph{causal chains of
  transitions} that are the \emph{building blocks} of macro-level
dynamics and information processing, as explained in detail below.

Another important feature of the DCM is its compact size.
While the dynamical landscape of an automata network, defined by its
state-transition graph (STG), grows exponentially with the number of
nodes -- $2^n$ in Boolean networks -- its DCM grows only linearly with
$2n$ units plus the number of t-units needed (which is the number of
schemata obtained from redescribing every automaton in the network):
$2n + \sum_{i=1}^n |F''_i|$.
Furthermore, the computation of a DCM is tractable even for very large
networks with thousands of nodes, provided the in-degree of these
nodes is not very large.
In our current implementation, we can exhaustively perform
schema redescription of automata with $k \leq k_{\textrm{max}} \approx
20$; that is, LUTs containing up to $2^{20}$ entries.
It is very rare that dynamical models of biochemical regulation have
molecular species that depend on more than twenty other variables (see
e.g. \cite{Thieffry:1998fk}).
Therefore, this method can be used to study canalization and
control in all discrete models of biochemical regulation we have
encountered in the literature, which we will analyse elsewhere.

It is important to emphasize that the integration of the CMs of
individual automata into the DCM does not change the control logic
encoded by each constituent CM, which is equivalent to the logic encoded in the original LUT (after removal of
redundancy).
Therefore, there is no danger of violating the logic encoded
in the original LUT of any automaton in a given BN.
However, it is necessary to ensure that any initial conditions
specified in the DCM do not violate the laws of contradiction and
excluded middle.
This means, for instance, that no initial condition of the DCM can
have the two (\emph{on} and \emph{off}) s-units for the same automaton
firing simultaneously.

\begin{figure}
\includegraphics{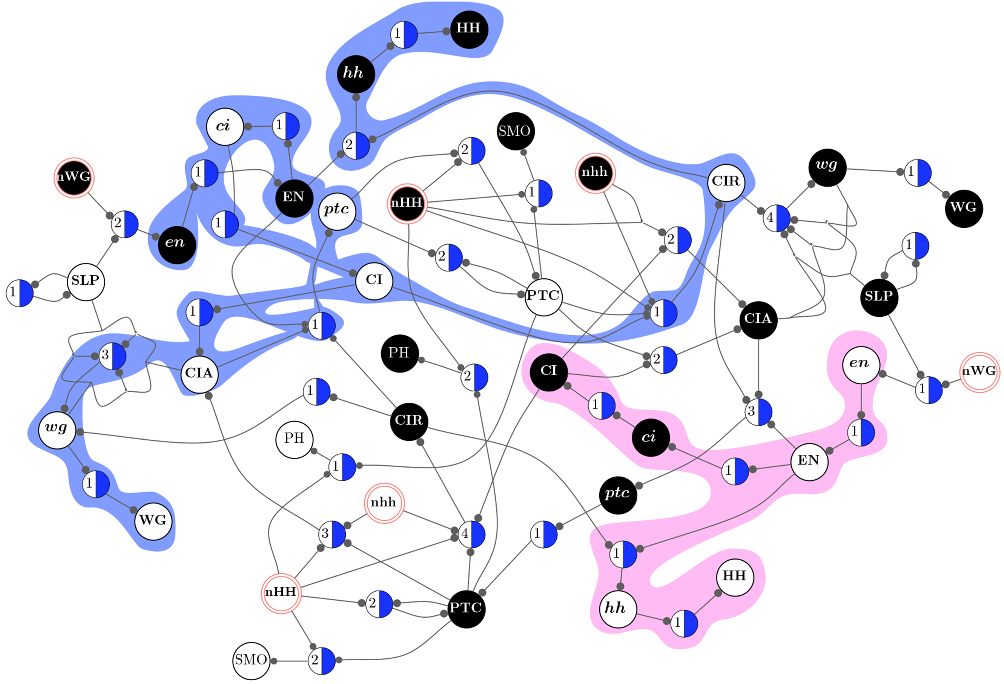}
\caption{{\bf Dynamics Canalization Map for a single cell of the SPN
    Model.} Also depicted are pathway modules $\mathcal{M}_1$ (pink) and
  $\mathcal{M}_2$ (blue), whose initial conditions depend exclusively on the
  expression and inhibition of input node SLP and of WG in neighbouring
  cells (the nWG spatial-signals). $\mathcal{M}_1 = \neg
  \textnormal{nWG} \vee \textnormal{SLP}$, $\mathcal{M}_2 = \neg
  \mathcal{M}_1$ (see details in text).}
\label{fig:spn_dcm}
\end{figure}

The DCM for a single cell in the SPN model is shown in
Figure~\ref{fig:spn_dcm}.
The spatial signals from adjacent cells are highlighted using units
with a double border $(nhh, n\textrm{HH and $n$WG})$.
For the simulations of the spatial SPN model described in
subsequent sections, we use four coupled single-cell DCMs (each as in
Figure~\ref{fig:spn_dcm}) to represent the dynamics of the four-cell
parasegment, where nodes that enable inter-cellular
regulatory interactions are appropriately linked, as defined in the
original model.
Also, as in the original model, we assume periodic boundary conditions
for the four-cell parasegment: the posterior cell is adjacent to the
anterior cell.
When making inferences using the DCM, we use \emph{signal} to refer to
the firing of a s-unit and the transmission of this information
through its output fibres.
When a s-unit fires in the DCM, it means that its corresponding
automaton node in the original BN transitioned to the state
represented by the s-unit.
We also use \emph{pathway} to refer to a logical sequence of signals
in the DCM.
We highlight two \emph{pathway modules} in the DCM of the SPN in
Figure~\ref{fig:spn_dcm}: $\mathcal{M}_1$ and $\mathcal{M}_2$.
The first is a pathway initiated by either the inhibition of WG in
neighbour cells, or the expression of SLP upstream in the same cell.
That is, the initial pattern for this module is $\mathcal{M}_1^0 = \neg n\textnormal{WG} \vee \textnormal{SLP}$.
The initiating signal for  $\mathcal{M}_2$ is defined by the
negation of those that trigger the first: $\mathcal{M}_2^0 = \neg
\mathcal{M}_1^0 = n\textnormal{WG} \wedge \neg \textnormal{SLP}$.
Both modules follow from (external or upstream) input signals to a
single cell in the SPN; they do not depend at all on the initial
states of nodes (molecular species) of the SPN inside a given cell.
Yet, both of these very small set of initial signals necessarily cause
a cascade of other signals in the network over time.
$\mathcal{M}_1$ is the only pathway that leads to the inhibition of
$en$ (and EN) as well as the expression of $ci$ (and CI).
It also causes the inhibition of $hh$ and HH, both of which function
as inter-cellular signals for adjacent cells -- this inhibition can be
alternatively controlled by the expression of CIR, which is not part
of neither $\mathcal{M}_1$ nor $\mathcal{M}_2$.
Since $\mathcal{M}_1^0$ is a disjunction, its
terms are equivalent: either the inhibition of $n$WG or the upstream
expression of SLP control the same pathway, regardless of any other
signals in the network.
$\mathcal{M}_2$ is the only pathway that leads to the expression of
$en$ (and EN) as well as the inhibition of $ci$ (and CI);
It also causes the inhibition of CIA, $ptc$ and CIR -- these
inhibitions can be alternatively controlled by other pathways.
If the initial conditions $\mathcal{M}_2^0$ are sustained for long
enough (steady-state inputs), the downstream inhibition of CIA and
sustained inhibition of SLP lead to the inhibition of $wg$ (and WG);
likewise, from sustaining $\mathcal{M}_2^0$, the
downstream expression of EN and inhibition of CIR lead to the
expression of $hh$ (and HH).
Since $\mathcal{M}_2^0$ is a conjunction, both
terms are required: both the expression of $n$WG and the upstream
inhibition of SLP are necessary and sufficient to control this
pathway module, regardless of any other signals in the network.

$\mathcal{M}_1$ and $\mathcal{M}_2$ capture a cascade of
state transitions that are inexorable once their initiating signals ($\mathcal{M}_1^0$ and $\mathcal{M}_2^0$)
are observed:
$\mathcal{M}_1 = \{\neg en,$$ \neg \textnormal{EN}, $ $\neg
hh, $ $\neg $ $\textnormal{HH}, $ $ci,$ $ \textnormal{CI} \}$
and
$\mathcal{M}_2 = \{ \neg ci,$ $ \neg \textnormal{CI},$ $ \neg
\textnormal{CIA}, $ $\neg wg, $ $ \neg \textnormal{WG}, $ $ \neg
\textnormal{CIR}, $ $\neg ptc, $ $ en, $ $\textnormal{EN}, $ $ hh, $ $\textnormal{HH} \}$.
Furthermore, these cascades are \emph{independent} from the states
of other nodes in the network.
As a consequence, the transitions within a module are insensitive to
delays once its initial conditions are set (and maintained in the case
of $\mathcal{M}_2$ as shown).
The \emph{dynamics} within these portions of the DCM can thus be seen
as \emph{modular}; these pathway modules can be \emph{decoupled} from
the remaining regulatory dynamics, in the sense that they are not
affected by the states of any other nodes other than their initial
conditions.
Modularity in complex networks has been typically defined as sub-graphs
with high intra-connectivity\cite{Fortunato:2009ly}.
But such structural notion of community structure does not capture the
dynamically decoupled behaviour of pathway modules such as
$\mathcal{M}_1$ and $\mathcal{M}_2$ in the SPN.
Indeed, it has been recently emphasized that understanding modularity
in complex molecular networks requires accounting for dynamics
\cite{Alexander:2009uq}, and new measures of modularity in
multivariate dynamical systems have been proposed by our group
\cite{Kolchinsky:2011fk}.
We will describe methods for automatic detection of dynamical
modularity in DCMs elsewhere.

Collective computation in the macro-level dynamics of
automata networks ultimately relies on the interaction of these
pathway modules.
Information gets integrated as modules interact with one another, in
such a way that the timing of module activity can have an effect on
downstream transitions.
For instance, the expression of CI via $\mathcal{M}_1$ can
subsequently lead to the expression of CIA, provided that $nhh$ is
expressed -- and this is controlled by $\mathcal{M}_2$ in the adjacent
cells.
The expression of CI can also be seen as a necessary initial condition
to the only pathway that results in the expression of CIR, which also
depends on the inhibition of $nhh$ and $n$HH and the expression of
PTC, which in turn depends on the interaction of other modules, and so on.
As these examples show, pathway modules allow us to uncover the
building blocks of macro-level control -- the collective computation
of automata network models of biochemical regulation.
We can use them, for instance, to infer which components exert most
control on a target collective behaviour of interest, such as the
wild-type expression pattern in the SPN.
Indeed, modules $\mathcal{M}_1$ and $\mathcal{M}_2$ in the SPN model,
which include a large proportion of nodes in the DCM, highlight how
much SLP and the spatial signals from neighbouring cells control the
dynamical behaviour of segment polarity gene regulation in each
individual cell.
Particularly, they almost entirely control the expression and
inhibition of EN and WG; as discussed further below.
The behaviour of these proteins across a four-cell parasegment mostly
define the attractors of the model (including wild-type).
The transitions of intra-cellular nodes are thus more controlled by
the states of `external' nodes than by the initial pattern of
expression of genes and proteins in the cell itself.
This emphasizes the well-known spatial constraints imposed on each
cell of the fruit fly's developmental system
\cite{Zallen:2004wg,Lu:2010wn}.
We next study and quantify this control in greater detail.

\subsubsection*{Dynamical unfolding}
%\addcontentsline{toc}{subsubsection}{Dynamical unfolding}

A key advantage of the DCM is that it allows us to study the behaviour of
the underlying automata network without the need to specify the state
of all of its nodes.
Modules $\mathcal{M}_1$ and $\mathcal{M}_2$ are an example of how the
control that a very small subset of nodes exerts on the dynamics of
SPN can be studied.
This can be done because, given the schema redescription that defines
the DCM, subsets of nodes can be assumed to be in an \emph{unknown}
state.
Since the schema redescription of every automaton in the DCM is
\emph{minimal} and \emph{complete} (see micro-level canalization
section), every possible transition that can occur is accounted for in
the DCM.
By implementing the DCM as a threshold network, we gain the ability to
study the dynamics of the original BN by setting the states of subsets
of nodes. This allows us study convergence to attractors, or other
patterns of interest, from knowing just a few nodes.

More formally, we refer to an initial pattern of interest of a BN
$\mathcal{B}$ as a \emph{partial configuration}, and denote it by
$\bm{\hat{x}}$.
For example,  $\mathcal{M}_1^0$ is a
partial configuration $\bm{\hat{x}_1} = \mathcal{M}_1^0 = SLP \vee \neg nWG$, where the
states of all other nodes is $\#$, or unknown.
We refer to \emph{dynamical unfolding} as the sequence of transitions
that necessarily occur after an initial partial configuration $\bm{\hat{x}}$,
and denote it by $\sigma(\bm{\hat{x}}) \leadsto \bm{\mathcal{P}}$, where
$\bm{\mathcal{P}}$ is an \emph{outcome pattern} or configuration.
From the DCM of the single-cell SPN model (Figure~\ref{fig:spn_dcm}), we have $\sigma(\mathcal{M}_1^0) \leadsto
\mathcal{M}_1$ and $\sigma(\mathcal{M}_2^0) \leadsto
\mathcal{M}_2$.
An outcome pattern can be a fully specified attractor $\mathcal{A}$, but it can also
be a partial configuration of an attractor where some nodes remain
unknown -- for instance, to study what determines the states of a specific subset of nodes of interest in the network.
In the first case, it can be said that $\bm{\hat{x}}$ \emph{fully controls} the network dynamics towards attractor $\mathcal{A}$.
In the second, control is exerted only on the subset of nodes with
determined logical states.

The ability to compute the dynamical unfolding of a BN from partial
configurations is a key benefit of the methodology introduced here: it
allows us to determine how much partial configurations of
interest \emph{control} the collective dynamics of the network.
For instance, in the SPN model it is possible to investigate how much
the input nodes to the regulatory network of each cell control its
dynamics.
Or, conversely, how much the initial configuration of the
intra-cellular regulatory network is irrelevant to determining its
attractor.
The nodes within each cell in a parasegment of the SPN are sensitive
to three inter-cellular (external) input signals: $n$WG, $nhh$ and
$n$HH, and one intra-cellular (upstream) input, SLP.
Given that the formation of parasegment boundaries in
\emph{D. melanogaster} is known to be tightly spatially constrained
\cite{Zallen:2004wg,Lu:2010wn}, it is relevant to investigate how
spatio-temporal control occurs in the SPN model.
We already studied the control power of SLP and $n$WG, which lead to
modules $\mathcal{M}_1$ and $\mathcal{M}_2$.
We now exhaustively study the dynamical unfolding of all possible
states of the intra- and inter-cellular input signals.

\begin{figure}
\includegraphics{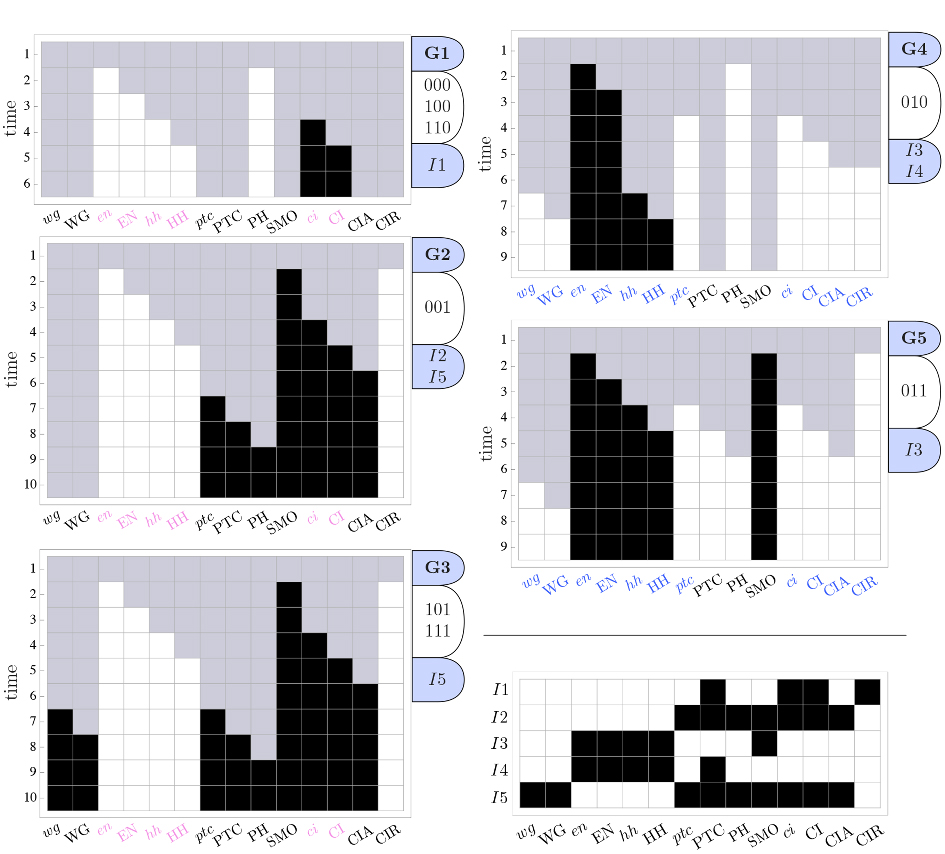}
\caption{{\bf Dynamical unfolding of the (single-cell) SPN with
    partial input configurations.}  The eight initial partial
  configurations tested correspond to the combinations of the
  steady-states of intra- and inter-cellular inputs SLP, $n$WG and
  $nhh$ (and where $n$HH and $nhh$ are merged into a single node,
  $nhh$). The specific state-combinations of these three variables is
  depicted on the middle (white) tab of each dynamical unfolding
  plot. Initial patterns that reach the same target pattern are
  grouped together in five groups $G1$ to $G5$ (identified in the top
  tab of each plot). The six input configurations in groups G1, G2,
  and G3 depict the dynamics where pathway module $\mathcal{M}_1$ is
  involved (nodes involved in this module are highlighted in pink.)
  The two input configurations in G4 and G5 depict the dynamics where
  pathway module $\mathcal{M}_2$ is involved (nodes involved in this
  module are highlighted in blue.) Three of the eight combinations are
  in $G1$ because they reach the same final configuration which,
  although partial, can only match the attractor $I1$. There are five
  possible attractor patterns of the SPN model for a single cell,
  shown in bottom right inset: $I1$ to $I5$ (see \S
  background). Attractors reached by each group are identified in the
  bottom tab of each plot. Groups $G2$ and $G4$ both unfold to an
  ambiguous target pattern that can end in $I2$ or $I5$ for $G2$, and
  $I3$ or $I4$ for $G4$. Finally, the initial partial configurations
  in groups $G3$ and $G5$ are sufficient to resolve the states of
  every node in the network.}
\label{fig:spn_dyn_unfold}
\end{figure}

We assume that SLP (upstream) and the (external) spatial signals are
in steady-state to study what happens in a single cell.
Since the state of $n$HH is the same as $nhh$ after
one time step, we consolidate those input signals into a single one:
$nhh$.
We are left with three input signals to the intra-cellular regulatory
network: nodes SLP, $n$WG and $nhh$.
Each of these three nodes can be in one of two states (\emph{on},
\emph{off}) and thus there are eight possible combinations of states for
these nodes.
Such simplification results in a
  non-spatial model and this was done previously by Willadsen \& Wiles
  \cite{Willadsen:2007hc}.
Setting each such combination as the initial partial configuration $\bm{\hat{x}}$,
and allowing the DCM to compute transitions, yields the results shown in Figure
\ref{fig:spn_dyn_unfold}.
We can see that only two of the outcome patterns reached by the eight
input partial configurations are ambiguous about which of the final
five possible attractors is reached. Each individual cell in a
  parasegment can only be in one of five attractor patterns $I1-I5$
  (see \S background).
This is the case of groups $G2$ and $G4$ in Figure
\ref{fig:spn_dyn_unfold}.
For all the other input partial configurations, the resulting outcome
pattern determines the final attractor.
We also found that for almost every input partial configuration, the
states of most of the remaining nodes are also resolved; in particular
the nodes that define the signature of the parasegment attractor --
Engrailed (EN) and Wingless (WG) -- settle into a defined
steady-state.
Notice also that for two of the input partial configurations (groups
$G3$ and $G5$ in Figure
\ref{fig:spn_dyn_unfold}), the states of every node in the network
settle into a fully defined steady-state.
The picture of dynamical unfolding from the intra- and inter-cellular
inputs of the single-cell SPN network also allows us to see the roles
played by modules $\mathcal{M}_1$ and $\mathcal{M}_2$ in the
dynamics.
The six input configurations in groups G1, G2, and G3 depict the
dynamics where $\mathcal{M}_1$ is involved, while the two input
configurations in G4 and G5 refer to $\mathcal{M}_2$ (node-states of
each module in these groups appear shaded in Figure
\ref{fig:spn_dyn_unfold}).
By comparing the resulting dynamics, we can see clearly the effect of
the additional information provided by knowing if $nhh$ is expressed
or inhibited; we also see that the dynamics of the modules is
unaffected by other nodes, as expected.

It is clear from these results that (single-cell) cellular dynamics in the SPN is
almost entirely controlled from the inputs alone.
We can say that extensive micro-level canalization leads the
macro-level network dynamics to be highly canalized by external inputs
-- a point we explore in more detail below.
For the dynamical unfolding depicted in Figure
\ref{fig:spn_dyn_unfold} we assumed that the three input signals to
the intra-cellular regulatory network are in steady-state, focusing on
a single cell.
This is not entirely reasonable since inter-cellular signals are regulated
by spatio-temporal regulatory dynamics in the full spatial SPN model.
We thus now pursue the identification of \emph{minimal} partial
configurations that guarantee convergence to outcome patterns of
interest in the spatial SPN model, such as specific (parasegment) attractors.

\subsubsection*{Minimal configurations}
%\addcontentsline{toc}{subsubsection}{MCs}

To automate the search of minimal configurations that converge to
patterns of interest, we rely again on the notion of schema
redescription, but this time for network-wide configurations rather
than for individual automata LUTs.
Notice that the eight input partial configurations used in the
dynamical unfolding scenarios described in Figure
\ref{fig:spn_dyn_unfold} are wildcard schemata of network
configurations: the state of the 14 inner nodes is \emph{unknown}
(wildcard), and only three (input) nodes (SLP, nWG,$nhh$) are set to a
combination of Boolean states.
Each of these eight schemata redescribes $2^{14}$ possible
configurations of the single-cell SPN.
Six of the eight input schemata converge to one of the five possible attractors for
inner nodes in a single cell of the SPN model (Figure \ref{fig:spn_dyn_unfold}).
We can thus think of those six schemata as \emph{minimal
  configurations} (MCs) that guarantee convergence to patterns
(e.g. attractors) of interest.

More specifically, a MC is a 2-symbol schema $\bm{x}''$ that
redescribes a set of network configurations that converge to target
pattern $\bm{\mathcal{P}}$; when the MC is a wildcard schema, it is
denoted by $\bm{x}'$.
Therefore, $\sigma(\bm{x}'') \leadsto \bm{\mathcal{P}}$.
MC schemata, $\bm{x}''$ or $\bm{x}'$, are network configurations where
the truth value of each constituent automaton can be 0, 1, or $\#$
(unknown); symmetry groups are allowed for $\bm{x}''$ and identified
with position-free symbols $\circ_m$ (see Micro-level canalization
section).
An MC schema redescribes a subset $\Theta$ of the set of
configurations $\bm{X}$: $\Theta \equiv \{\bm{x} \in \bm{X}: \bm{x}
\rightarrowtail \bm{x}''\}$.
A partial configuration is a MC if no Boolean state in it can be raised to the unknown state ($\#$) and still
guarantee that the resulting partial configuration converges to
$\bm{\mathcal{P}}$.
In the case of a two-symbol schema, no group-invariant enput can be
enlarged (include additional node-states) and still guarantee
convergence to $\bm{\mathcal{P}}$.
Finally, the target pattern $\bm{\mathcal{P}}$ can be a specific
network configuration (e.g. an attractor), or it can be a set of
configurations of interest (e.g. when only some genes or proteins are expressed).
After redescription of a set of configurations $\bm{X}$ of a BN -- a
subset or its full dynamical landscape -- we obtain a set of
two-symbol MCs $\bm{X}''$; a set of wildcard MCs is denoted by
$\bm{X}'$.
Similarly to micro-level schemata, we can speak of enputs of MCs. In
this context, they refer to individual and sets of node-states in the
network that are essential to guarantee convergence to a target
pattern.

The dynamical unfolding example of the single-cell SPN model shows
that to converge to the attractor $I1$
(Figure~\ref{fig:spn_dyn_unfold}, G1), only the states of the three
input nodes need to be specified, in one of three possible Boolean
combinations: $000, 100$ or $110$ for the nodes SLP, $n$WG and $nhh$;
all other (inner) nodes may be unknown ($\#$).
Moreover, these three initial patterns can be further redescribed into
two schemata: $\bm{X}' = \{ \{\#, 0,0\}, \{1, \#,0\} \}$.
This shows that to guarantee converge to  $I1$, we
only need to know the state of two (input) nodes: either $n$WG $= nhh
= 0$, or SLP = 1 and $nhh = 0$. All other nodes in the single-cell
model can remain unknown. Therefore, the MCs for attractor pattern
$I1$ are:

%\begin{equation}
\begin{align}
\bm{X}' = \{ &\#\#\#\#\#\#\#\#\#\#\#\#\#\#\#00, \nonumber \\
             &\#\#\#\#\#\#\#\#\#\#\#\#\#\#1\#0 \}
\label{eq:MCs_SPN_I1}
\end{align}
%\end{equation}

\noindent where the order of the inner nodes is the same as in
Figure~\ref{fig:spn_dyn_unfold}, and the last three nodes are SLP,
$n$WG and $nhh$ in that order.
Notice that in this case there is no group-invariance, so $\bm{X}'' = \bm{X}'$.
Any initial configuration not redescribed by $\bm{X}'$, does not
converge to pattern $I1$.
Therefore, these MCs reveal the enputs (minimal set of node-states)
that \emph{control} network dynamics towards attractor $I1$: $nhh$
must remain unexpressed, and we must have either SLP expressed, or
$n$WG unexpressed.
However, as mentioned above, this example refers to the case when the
three input nodes are in steady-state.
For the single-cell SPN, the steady-state assumption is reasonable.
But for the spatial SPN, with parasegments of four cells, we cannot
be certain that the spatial signals ($n$WG and $nhh$) have
reached a steady-state at the start of the dynamics.
Therefore, we now introduce a procedure for obtaining MCs, without the
steady-state assumption, which we apply to the spatial SPN network
model.

It was discussed previously that individual automata in BN models of biochemical
regulation and signalling very rarely have large numbers of input
variables.
This allows tractable computation of two-symbol schema redescription
of their LUTs (see micro-level section).
In contrast, computing MCs for network configurations easily becomes more
computationally challenging.
Even for fairly small networks with $n \approx 20$, the size of their
dynamical landscape becomes too large to allow full enumeration of the
possible configurations and the transitions between them.
As shown above, it is possible to identify pathway modules, and to
compute dynamical unfolding on the DCM, without knowing the
STG of very large BNs, but it remains not feasible to fully redescribe
their entire dynamical landscape.

One way to deal with high-dimensional spaces is to
resort to \emph{stochastic search} (see e.g. \cite{Mitchell:1996ml}).
We use stochastic search to obtain MCs
that are guaranteed to converge to a pattern of interest
$\bm{\mathcal{P}}$.
We start with a \emph{seed} configuration known to converge to $\bm{\mathcal{P}}$.
Next, a random node in a Boolean state is picked, and
changed to the unknown state.
The resulting partial configuration is then allowed to unfold to
determine if it still converges to $\bm{\mathcal{P}}$.
If it does, the modified configuration becomes the new seed.
The process is repeated until no more nodes can be `raised' to the unknown state and still ensure convergence to
$\bm{\mathcal{P}}$.
Otherwise, the search continues picking other nodes.
The output of this algorithm (detailed in \emph{Supporting
 text S4}) is thus a single wildcard MC.
Afterwards, the goal is to search for \emph{sets} of MCs
that converge to $\bm{\mathcal{P}}$.
We do this in two steps: first we search for a set of MCs derived from
a single seed, followed by a search of the space of possible different
seeds that still converge to $\bm{\mathcal{P}}$.
We use two `tolerance' parameters to determine when to stop searching.
The first, $\delta$, specifies the number of times a single seed must
be `reused' in the first step.
When the algorithm has reused the seed $\delta$ consecutive times
without finding any new MCs, the first step of the MC search stops.
The second tolerance parameter, $\rho$, is used to specify when
to stop searching for new seeds from which to derive MCs.
When $\rho$ consecutively generated random (and different) seeds are
found to be already redescribed by the current set of MCs, the algorithm stops.
Both parameters are reset to zero every time a new MC is identified.
These two steps are explained in greater detail in 
\emph{Supporting text S4}.

The two-step stochastic search process results in a set of wildcard
schemata $\bm{X}'$ that redescribe a given set of configurations
$\bm{X}$ guaranteed to converge to pattern $\bm{\mathcal{P}}$.
We next obtain a set of two-symbol MCs $\bm{X}''$ from $\bm{X}'$, by
identifying group-invariant subsets of nodes using the same method
described in the micro-level canalization section.
Since $\bm{X}'$ can be quite large (see below), this computation
can become challenging.
In this case, we restrict the search for symmetric groups in $X'$ that
redescribe a minimum number $\beta$ of wildcard MCs $\bm{x}'$.

Notice that it is the DCM, implemented as a threshold network, that
allows us to pursue this stochastic search of MCs. With the original
BN, we cannot study dynamics without setting every automaton to a
specific Boolean truth value. With the DCM, obtained from micro-level
canalization, we are able to set nodes to the unknown state and study
the dynamical unfolding of a partial configuration (see previous
subsection) to establish convergence to a pattern of interest.
Therefore, the DCM helps us link micro-level canalization to
macro-level behaviour. Let us exemplify the approach with the SPN
model.

We started our study of MCs in the spatial SPN model, with a set of
\emph{seed} configurations $\bm{X}_{\textrm{bio}}$ that contains
the known initial configuration of the SPN (shown in Figure
\ref{fig:spn_parasegment}), the wild-type attractor (Figure
\ref{fig:spatial_attractors}a), and the five configurations in the
dynamic trajectory between them.
When searching for MCs using these seed configurations we set $\delta=10^5$.
This resulted in a set containing 90 wildcard MCs
$\bm{X}'_{\textrm{bio}}$ (available in \emph{Supporting data S7}).
Using the set $\bm{X}'_{\textrm{bio}}$, we performed the two-step
stochastic search with $\rho=10^6$ and $\delta=10^5$.
This resulted in a much larger set of 1745 wildcard MCs (available in
\emph{Supporting data S8}) which guarantee convergence to wild-type:
$\bm{X}'_{\textrm{wt}} \supset \bm{X}'_{\textrm{bio}}$.
The number of literal enputs in each MC contained in this set varies from 23
to 33 -- out of the total 60 nodes in a parasegment.
In other words, from all configurations in $\bm{X}_{\textrm{wt}}$ we
can ascertain that to guarantee convergence to the wild-type
attractor, we need only to control the state of a minimum of 23 and a
maximum of 33 of the 60 nodes in the network. %
Equivalently, 27 to 37 nodes are irrelevant in steering the dynamics
of the model to the wild-type attractor -- a high degree of
canalization we quantify below.

We chose to study two further subsets of $\bm{X}'_{\textrm{wt}}$
separately: $\bm{X}'_{\textrm{noP}}$ and $\bm{X}'_{\textrm{min}}$.
The first (available in \emph{Supporting data S9}) is the subset of MCs that do not
have enputs representing expressed (\emph{on}) proteins, except
SLP$_{3,4}$ -- since SLP in cells 3 and 4 is assumed to be present
from the start, as determined by the pair-rule gene family (see
\cite{Albert:2003ij} and introductory section).
This is a subset of interest because it corresponds to the expected
control of the SPN at the start of the segment-polarity dynamics,
including its known initial configuration (Figure
\ref{fig:spn_parasegment}); thus $ \bm{X}'_{\textrm{noP}} \subset \bm{X}'_{\textrm{wt}}$.
The second, $\bm{X}'_{\textrm{min}} \subset \bm{X}'_{\textrm{wt}}$ is
the subset of MCs with the smallest number of enputs (available in
\emph{Supporting data S10}.
This corresponds to the set of 32 MCs in $\bm{X}'_{\textrm{wt}}$ that
have only 23 enputs each.
This is a subset of interest because it allows us to study how the
unfolding to wild-type can be guaranteed with the smallest possible
number of enputs.
Notice that $\bm{X}'_{\textrm{min}}$ redescribes a large subset of
configurations in $\bm{X}_{\textrm{wt}}$ because it contains the MCs
with most redundant number of nodes.
These sets of wildcard MCs are available in \emph{Supporting
  data S7,S8, S9} and \emph{S10}; Table \ref{tab:MC_Summary} contains
their size.

There are severe computational limitations to counting exactly the
number of configurations redescribed by each set of MCs, since it
depends on using the inclusion/exclusion principle
\cite{Bjorklund:2006uq} to count the elements of intersecting sets
(MCs redescribe overlapping sets of configurations).
See \emph{Supporting text S6} for further details.
We can report the exact value for $|\bm{X}_{\textrm{noP}}| = 8.35
\times 10^{10}$, which is about $14\%$ of the number of configurations
-- or pre-patterns -- estimated by Albert \& Othmer
\cite{Albert:2003ij} to converge to the wild-type attractor $(6 \times
10^{11})$.
Using the inclusion/exclusion principle, it was also computationally
feasible to count the configurations redescribed by a sample of 20 of
the 32 MCs in $\bm{X}'_{\textrm{min}}: 9.6 \times 10^{11}$.
Since this sample of 20 MCs is a subset of $\bm{X}'_{\textrm{min}}$,
which is a subset of $\bm{X}'_{\textrm{wt}}$, we thus demonstrate that
$|\bm{X}_{\textrm{wt}}| \geq |\bm{X}_{\textrm{min}}| \geq 9.6 \times
10^{11}$, which is $1.6$ times larger than the previously estimated
number of pre-patterns converging to the wild-type attractor
\cite{Albert:2003ij}.
This means that the wild-type attraction basin is considerably (at least 1.6 times) larger than
previously estimated, with a lower bound of at least $9.6 \times 10^{11}$ network configurations.
Although it was not computationally feasible to provide exact counts
for the remaining MC sets, it is reasonable to conclude that the set
$\bm{X}'_{\textrm{wt}}$ redescribes a significant proportion of the
wild-type attractor basin, given the number of configurations redescribed by 20 of its most canalized MCs in comparison to the previous estimate of its size.
Indeed, we pursued a very wide stochastic search with large tolerance
parameters, arriving at a large number (1745) MCs, each of which
redescribes a very large set of configurations.
For instance, each MC with the smallest number of enputs (23) alone
redescribes $1.37 \times 10^{11}$ configurations, which is about
$23\%$ of the original estimated size of the wild-type attractor basin,
and $14\%$ of the lower bound for the size of the attractor basin we computed above.
Given the large number of MCs in the $\bm{X}'_{\textrm{wt}}$ set, even
with likely large overlaps of configurations, much of the attractor
basin ought to be redescribed by this set.

From $\bm{X}'_{\textrm{wt}}$, we derived two-symbol MC sets using $\beta=8$.
That is, due to the computational limitations discussed previously, we
restricted the search to only those two-symbol MCs $\bm{x}''$ that
redescribe at least $\beta=8$ wildcard MCs $\bm{x}'$.
Given that configurations of the spatial SPN are defined by $60$
automata states, the group-invariance enputs we may have missed with
this constraint are rather trivial.
For instance, we may have missed MCs with a single group-invariant
enput of 3 variables (any group-invariant enput with 4 variables would
be found), or MCs with 2 distinct group-invariant enputs of 2
variables each (any MCs with 3 group-invariant enputs would be found.)
With this constraint on the search for two-symbol MCs, we
identified only the pair of two-symbol MCs depicted in Figure
\ref{fig:min_two_sym}: $\{\bm{x}''_1, \bm{x}''_2 \}$ -- each
redescribing 16 wildcard MCs -- the MCs redescribed are available in
\emph{Supporting data S13}.
These two MCs redescribe $1.95 \times 10^{11}$ configurations; that is,
about $32\%$ of the wild-type attraction basin as estimated by
\cite{Albert:2003ij},
or $20\%$ of the lower bound for the size of the attractor basin we
computed above -- a very substantial subset of the wild-type attractor
basin.

No other two-symbol MCs redescribing at least eight wildcard MCs were
found in the set $\bm{X}'_{\textrm{wt}}$. Therefore, $\bm{X}''_{\textrm{wt}}$ is comprised of
the wildcard MCs in $\bm{X}'_{\textrm{wt}}$ with the addition of
$\{\bm{x}''_1, \bm{x}''_2 \}$ and removal of the wildcard MCs these
two schemata redescribe. Table \ref{tab:MC_Summary} contains the size
of all MC sets.
Moreover, $\{\bm{x}''_1, \bm{x}''_2 \}$ have no intersecting schemata
with the additional three subsets of $\bm{X}''_{\textrm{wt}}$
we studied.
This means that the two-symbol
redescription (with $\beta=8$) is equal to the wildcard redescription of the sets of configurations
$\bm{X}_{\textrm{bio}}$, $\bm{X}_{\textrm{noP}}$ and
$\bm{X}_{\textrm{min}}$.
The pair of two-symbol MCs identified denote two very similar minimal
patterns that guarantee convergence to the wild-type attractor.
In both MCs, the pairs of nodes $wg_{2,4}$, HH$_{2,4}$ as well as
$ci_4$ and CI$_4$ are marked with distinct position-free symbols.
In other words, they have three identical group-invariant enputs.
For $\bm{x}''_1$ a fourth group-invariant enput comprises the nodes
$hh_{1,3}$, while for $\bm{x}''_2$ the fourth group-invariant enput
contains the nodes HH$_{1,3}$.
For $\bm{x}''_2$ there is an extra literal enput: $ptc_4 = 0$ ($ptc$
gene in fourth cell is unexpressed). The remaining literal enputs are
identical to those of $\bm{x}''_1$.
The group-invariance in these MCs is not very surprising considering
the equivalent roles of neighbouring hedgehog and Wingless for
intra-cellular dynamics -- as discussed previously when the SPN's DCM
was analysed. Notice that most group-invariance occurs for the same
genes or proteins in alternative cells of the parasegment; for
instance, $wg$ expressed in either cell 2 or cell 4.
Nonetheless, both two-symbol MCs offer two minimal conditions to
guarantee convergence to the wild-type attractor, which includes a
very large proportion of the wild-type attractor basin. Therefore,
they serve as a parsimonious prescription for analysts who wish to
control the macro-level behaviour (i.e. attractor behaviour) of this
system.
Finally, the MCs obtained observe substantial macro-level canalization
which we quantify below.

\subsection*{Quantifying Macro-level canalization}
%\addcontentsline{toc}{subsection}{Quantifying Macro-level canalization}

In the micro-level canalization section, we defined measures of
\emph{input redundancy}, \emph{effective connectivity} and \emph{input
  symmetry} to quantify micro-level canalization from the schema
redescription of individual automata.
Since we can also redescribe configurations that produce network
dynamics, leading to the minimal configurations (MCs) of the previous
section, we can use very similar measures to quantify macro-level
canalization and control.
At the macro-level, high canalization means that network dynamics are
more easily controllable: MCs contain fewer necessary and sufficient
node-states (enputs) to guarantee convergence to an attractor or
target pattern $\bm{\mathcal{P}}$.
Similarly to the micro-level case, we first define upper and lower
bounds of \emph{node redundancy} computed from the set of MCs
$\bm{X}''$ for a target pattern:

\begin{equation}
\bar{n}_{\textrm{r}}(\bm{X},\bm{\mathcal{P}}) = \frac{\displaystyle \sum_{\bm{x} \in \bm{X}} \max_{ {\theta : \bm{x} \in \Theta_\theta}}\left(n_\theta^\# \right)}{|\bm{X}|}
\label{upper_node_red}
\end{equation}

\begin{equation}
\underline{n}_{\textrm{r}}(\bm{X}, \bm{\mathcal{P}}) = \frac{\displaystyle \sum_{\bm{x} \in \bm{X}} \min_{ {\theta : \bm{x} \in \Theta_\theta}}\left(n_\theta^\# \right)}{|\bm{X}|}
\label{lower_node_red}
\end{equation}

These expressions tally the mean number of irrelevant nodes in
controlling network dynamics towards $\bm{\mathcal{P}}$ for all
configurations $\bm{x}$ of a set of configurations of interest
$\bm{X}$ (e.g. a basin of attraction).
The number of irrelevant nodes in a given MC $\bm{x}_{\theta}''$ is
the number of its wildcards $n_{\theta}^{\#}$.
Because each configuration $\bm{x}$ is redescribed by one or more MCs,
there are various ways to compute a characteristic number of
irrelevant nodes associated with the configurations, which is
nonetheless bounded by the maximum and minimum number of wildcards in
the set of MCs that redescribe $\bm{x}$.
Therefore, the expressions above identify all MCs whose set of
redescribed configurations $\Theta_{\theta}$ includes $\bm{x}$.
The upper (lower) bound of node redundancy, Equation
\ref{upper_node_red} (Equation \ref{lower_node_red}), corresponds to
considering the maximum (minimum) number of irrelevant nodes found for
all MCs that redescribe configuration $\bm{x}$ of
the interest set -- an optimist (pessimist) quantification of this
type of macro-level canalization.
Here we use solely the upper bound, which we refer to
henceforth simply as \emph{node redundancy} with the notation
$n_{\textrm{r}}(\bm{X}, \bm{\mathcal{P}})$.
Similarly to the micro-level case, the assumption is that the most
redundant MCs are always accessible for control of the network towards
pattern $\bm{\mathcal{P}}$.
The range for node redundancy is $ 0 \le
n_{\textrm{r}} \le n$, where $n$ is the number of nodes in the network.
When $n_{\textrm{r}}(\bm{X}, \bm{\mathcal{P}}) = n$ we have full node
irrelevance, or maximum canalization, which occurs only in the case of
networks where the state of every node is not dependent on
any input (that is, when $k_{\textrm{r}} = k$ for every node).
If $n_{\textrm{r}}(\bm{X}, \bm{\mathcal{P}}) = 0$, the state of every
node is always needed to determine convergence to $\bm{\mathcal{P}}$
and we have no macro-level canalization.

If some nodes of a network are irrelevant to steer dynamics to
$\bm{\mathcal{P}}$, from a control logic perspective, we can say that
$\bm{\mathcal{P}}$ is effectively controlled by a subset of nodes of
the network with fewer than $n$ nodes.
In other words, by integrating the micro-level control logic of
automata in a network into the DCM, we are able to compute MCs and
infer from those the macro-level \emph{effective control}, which is
not apparent from looking at connectivity structure alone:

\begin{equation}
n_{\textrm{e}}(\bm{X}, \bm{\mathcal{P}}) = n - n_{\textrm{r}}(\bm{X}, \bm{\mathcal{P}})
\label{lower_eff_n}
\end{equation}

\noindent whose range is $ 0 \le n_{\textrm{e}} \le n$. If
$n_{\textrm{e}}(\bm{X}, \bm{\mathcal{P}}) = 0$ it means full node
irrelevance, or maximum canalization.
When $n_{\textrm{e}}(\bm{X}, \bm{\mathcal{P}}) = n$, it means no
canalization i.e. one needs to control all $n$ nodes to guarantee
converge to $\bm{\mathcal{P}}$.

Macro-level canalization can also manifest \emph{alternative} control mechanisms.
The two-symbol schema redescription allows us to measure this form of
control by computing the mean number of nodes that participate in
group-invariant enputs, easily tallied by the number of position-free
symbols ($n_\theta^\circ$) in MC schemata $\bm{x}_{\theta}''$ that
characterize convergence to target pattern $\bm{\mathcal{P}}$.
Thus, we quantify the upper and lower bounds of \emph{node symmetry}
in a set of configurations of interest $\bm{X}$ related to target
pattern $\bm{\mathcal{P}}$ (e.g. a basin of attraction).

\begin{equation}
\bar{n}_{\textrm{s}}(\bm{X}, \bm{\mathcal{P}}) = \frac{  \displaystyle \sum_{\bm{x} \in \bm{X}}
 \max_{ {\theta : \bm{x} \in \Theta_\theta}}\left( n_\theta^\circ \right) }{|\bm{X}|}
\label{n_sym_upper}
\end{equation}

\begin{equation}
\underline{n}_{\textrm{s}}(\bm{X}, \bm{\mathcal{P}}) = \frac{  \displaystyle \sum_{\bm{x} \in \bm{X}}
 \min_{ {\theta : \bm{x} \in \Theta_\theta}}\left( n_\theta^\circ \right) }{|\bm{X}|}
\label{n_sym_lower}
\end{equation}

Here we use solely the upper bound, which we refer to henceforth
simply as node symmetry and denote by $n_{\textrm{s}}(\bm{X},
\bm{\mathcal{P}})$; its range is $[0, n]$.
Again, the assumption is that the most canalized MCs are always
accessible for control of the network towards pattern
$\bm{\mathcal{P}}$.
High (low) values mean that permutations of node-states are likely
(unlikely) to leave the transition unchanged.

Macro-level canalization in network dynamics is then quantified by two
types of redundancy: node redundancy (or its counterpart, effective
control) and node symmetry.
To be able to compare macro-level control in automata networks of
different sizes, we can compute \emph{relative} measures of
canalization:

\begin{equation}
n_{\textrm{r}}^{*}(\bm{X}, \bm{\mathcal{P}}) = \frac{n_{\textrm{r}}(\bm{X}, \bm{\mathcal{P}})}{n}; \quad n_{\textrm{e}}^{*}(\bm{X}, \bm{\mathcal{P}}) = \frac{n_{\textrm{e}}(\bm{X}, \bm{\mathcal{P}})}{n}; \quad n_{\textrm{s}}^{*}(\bm{X}, \bm{\mathcal{P}}) = \frac{n_{\textrm{s}}(\bm{X}, \bm{\mathcal{P}})}{n}
\label{relative__macro_canalization_measures}
\end{equation}

\noindent whose range is $[0, 1].$
Network dynamics towards a pattern of interest $\bm{\mathcal{P}}$ can
have different amounts of each form of canalization, which allows us
to consider four broad classes of control in network dynamics -- just
like the micro-level canalization case (see above).

The two MCs identified above for the single-cell SPN model
(Eq. \ref{eq:MCs_SPN_I1}), redescribe the full set of configurations
that converge to $I1$.
Since these MC schemata do not have group-invariant enputs, node
symmetry does not exist: $ {n}_{\textrm{s}}(\bm{X}, I1) = 0$.
Node redundancy and effective control is ${n}_{\textrm{r}}(\bm{X}, I1)
= 15$ and ${n}_{\textrm{e}}(\bm{X}, I1) = 2$, respectively.
In other words, even though the network of the single-cell SPN model
comprises $n=17$ nodes, to control its dynamics towards attractor
$I1$, it is sufficient to ensure that the states of only two nodes
remain fixed; the initial state of the other 15 nodes is irrelevant.
More concretely, $nhh$ must remain \emph{off} and
either SLP remains \emph{on} or $nwg$ remains
\emph{off}.
The relative measures become: $n_{\textrm{r}}^{*}(\bm{X}, I1) = 15/17$
($\approx 88\%$ of nodes are redundant to guarantee convergence to
attractor $I1$) $n_{\textrm{e}}^{*}(\bm{X}, I1) = 2/17$ (one only
needs to control $\approx 12\%$ of nodes to guarantee convergence to
attractor $I1$),
and $n_{\textrm{s}}^{*}(\bm{X}, I1) = 0$ (there is no node symmetry in these MCs).
This means that there is a large amount of macro-level canalization of
the node redundancy type -- and thus higher controllability -- in the
basins of attraction of the SPN model where pattern $I1$ is present.

The macro-level canalization measures above assume that the interest
set of configurations $\bm{X}$ can be enumerated.
Moreover, schema redescription of network configurations itself
assumes that $\bm{X}$ can be sufficiently sampled with our stochastic
search method (see previous sub-section).
The node symmetry measure additionally assumes that the set of
wildcard MCs obtained by stochastic search is not too large to compute
symmetric groups.
While these assumptions are easily met for micro-level analysis,
because LUT entries of individual automata in models of biochemical
regulation do not have very large number of inputs, they are more
challenging at the macro-level.
Certainly, canalization in the single-cell SPN model can be fully
studied at both the micro- and macro-levels -- see Figures
\ref{fig:spn_schemata} and \ref{fig:spn_canalization_qualitative} for
the former as well as example above for the latter.
But quantification of macro-level canalization of larger networks, such as
the spatial SPN model, needs to be estimated.
Therefore, in formulae \ref{upper_node_red}, \ref{lower_node_red},
\ref{n_sym_upper}, and \ref{n_sym_lower}, the set of
configurations $\bm{X}$ is sampled: $\bm{\hat{X}}$.
Configurations for $\bm{\hat{X}}$ are sampled from each MC in the set
$\bm{X}''$, proportionally to the number of configurations redescribed
by each MC -- i.e. roulette wheel sampling.
Configurations from a selected MC are sampled by ascribing Boolean
truth values to every wildcard in the MC schema; the proportion of
each of the truth values is sampled from a uniform distribution.
If a selected MC is a 2-symbol schema, the truth-values of
group-invariant enputs are also sampled from a uniform distribution of
all possible possibilities.
Naturally, the same configuration $\bm{x}$ can be
redescribed by more than one MC $\theta$.
In summary, macro-level canalization for larger networks is
quantified with the estimated measures: $\hat{n}_{\textrm{r}}$,
$\hat{n}_{\textrm{e}}$, and $\hat{n}_{\textrm{s}}$, as well as their
relative versions.

Tables \ref{tab:MC_Summary} and \ref{tab:MC_Summary2} summarize the
quantification of macro-level canalization estimated for the four MC
sets obtained above: $\bm{X}''_{\textrm{wt}}$,
$\bm{X}''_{\textrm{min}}$, $\bm{X}''_{\textrm{bio}}$, and
$\bm{X}''_{\textrm{noP}}$.
Effective control ($n_{\textrm{e}}$) ranges between $23$ and $26.2$ nodes
(out of $60$) for the four sets of MCs; this means (see
$n_{\textrm{e}}^{*}$) that only $38$ to $44\%$ of nodes need to be
controlled to guarantee convergence to wild-type.
This shows that there is substantial macro-level canalization in the
wild-type attractor basin; from $n_{\textrm{r}}^{*}$, we can see that
$56$ to $62\%$ of nodes are, on average, redundant to guarantee
convergence to wild-type.
On the other hand, macro-level canalization in the form of alternative
(or symmetric) control mechanisms is not very relevant on this
attractor basin, as observed by the low values of $n_{\textrm{s}}$ and
$n_{\textrm{s}}^{*}$: in the wild-type attractor basin, on average,
only approximately 1 out 60 nodes, or $1.6\%$ can permute.

\subsection*{Enput power and critical nodes}
%\addcontentsline{toc}{subsection}{Enput Power}

Every MC is a schema, and hence comprises a unique set of enputs, not
entirely redescribed by any other MC.
As defined in the micro-level canalization section, an enput $e$ can
be  literal -- a single node in a specific Boolean state -- or a
group-invariant enput: a set of nodes with a symmetry constraint.
Every enput $e$ in a given MC is essential to ensure convergence to a
pattern $\bm{\mathcal{P}}$, e.g. an attractor $\mathcal{A}$.
Consequently, if the state or constraint of $e$ is disrupted in the
MC, without gaining additional knowledge about the configuration of
the network, we cannot guarantee convergence to $\bm{\mathcal{P}}$.
How \emph{critical} is $e$ in a set of configurations $\bm{X}$
redescribed by an MC set $\bm{X}''$ -- such as the set of MCs that
redescribe a basin of attraction?
Since there are usually
alternative MCs that redescribe the possible dynamic trajectories to
$\bm{\mathcal{P}}$, the more $e$ appears in $\bm{X}''$, the more critical
it is in guaranteeing convergence to $\bm{\mathcal{P}}$.

\begin{figure}
\includegraphics{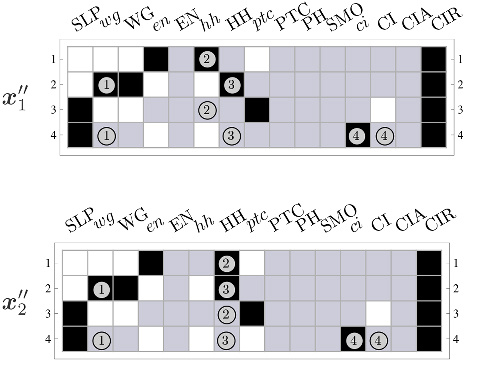}
\caption{{\bf Two-Symbol schemata with largest number of
    position-free symbols, obtained from redescription of
    $\bm{X}_{\textrm{wt}}$.} The pair $\{\bm{x}''_1, \bm{x}''_1 \}$
  were the two-symbol schemata obtained in our stochastic search; both
  include 4 pairs of symmetric node-pairs, each denoted by a circle
  and a numerical index. }
\label{fig:min_two_sym}
\end{figure}

For instance, in the two MCs shown in Equation \ref{eq:MCs_SPN_I1}, the
enput $e \equiv (nhh = 0)$ is common to both.
Therefore, disrupting it, without gaining additional knowledge about
the state of other nodes, would no longer guarantee convergence to the
attractor pattern $I1$ in the single-cell SPN dynamics.
Similarly, for the two-symbol MC set of the spatial SPN model, shown
in Figure \ref{fig:min_two_sym}, enputs $e \equiv (hh_{2,4} = 0)$ and
group-invariant enput $e \equiv (wg_2 = 1 \vee wg_4 = 1)$ appear in
both MCs.
Disrupting them, would no longer guarantee convergence to wild-type
attractor in the spatial SPN dynamics.

Let us quantify the potential disruption of target dynamics by
perturbation of enputs in an MC set.
The \emph{power} of an enput $e$ in a set of configurations $\bm{X}
\rightarrowtail \bm{X}'' : \sigma(\bm{x}) \leadsto \bm{\mathcal{P}}, \forall \bm{x} \in \bm{X}$,
is given by:

\begin{equation}
  \epsilon(e,\bm{X}'',\bm{\mathcal{P}}) =
  \frac{|\bm{X}_e|}{|\bm{X}|}
\label{eq:dominance}
\end{equation}

\noindent where $\bm{X}_e \subseteq \bm{X}$ is the subset of
configurations redescribed by $\bm{X}''$  that contain enput $e$:
$\bm{X}_e \equiv \{\bm{x} \in \bm{X}: \bm{x} \rightarrowtail
\bm{x}'' \wedge e \in \bm{x}'' \}$.
Thus, this measure yields the proportion of configurations in $\bm{X}$
redescribed by the MCs in which $e$ is an enput; its range is
$[0,1]$.
If an enput appears in every MC, as in the examples above, then
$\epsilon = 1$ -- in which case $e$ is said to have \emph{full power}
over $\bm{X}''$.
For the analysis of the SPN model below when $ 0.5 \le \epsilon < 1$, $e$ is a \emph{high power}
enput, when $ 0 < e < 0.5$ it is a \emph{low power} enput, and
when $\epsilon =0$ it is a \emph{null power} enput.
The larger the power of $e$, the more its perturbation is likely to
disrupt convergence to the target pattern $\bm{\mathcal{P}}$.
When $\bm{X}$ is too large, we estimate $\hat{\epsilon}$ -- similarly
to the canalization measures discussed in the previous subsection.

\begin{figure}
\includegraphics{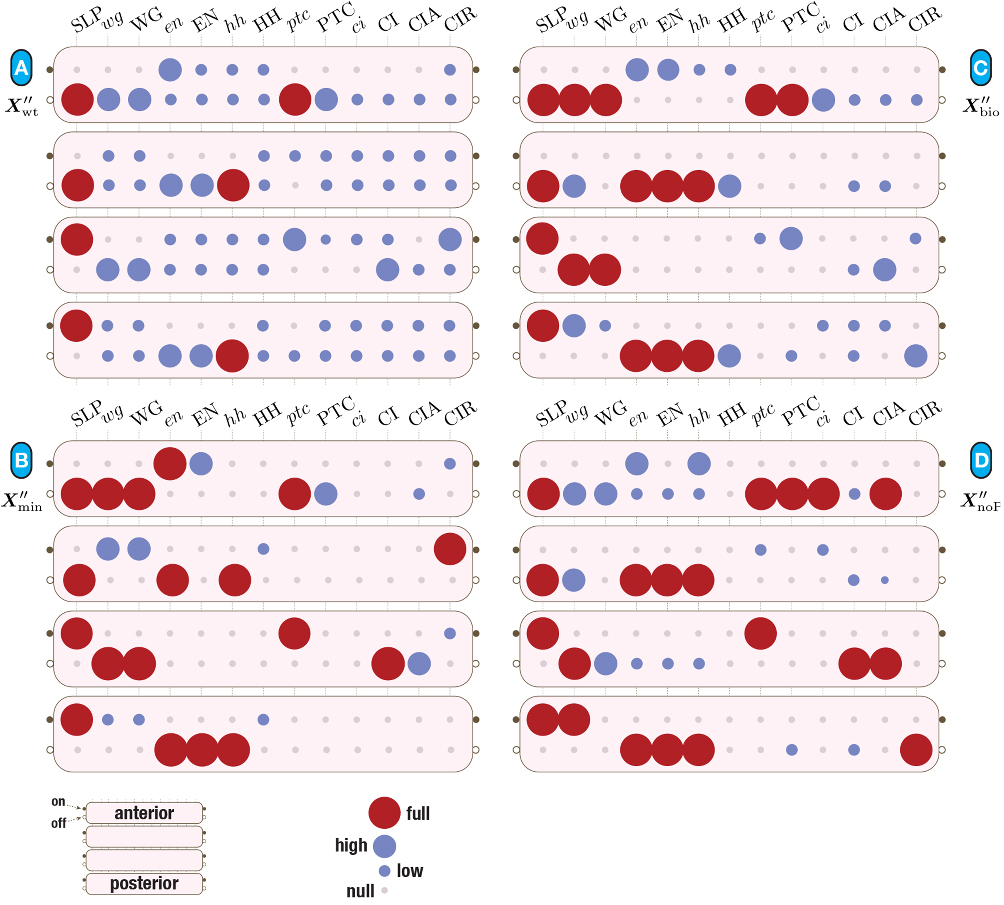}
\caption{{\bf Enput power in the wild-type basin of attraction of
    the spatial SPN model.} Enput power is shown for each of the four
  sets of MCs considered in our analysis: (A)
  $\bm{X}''_{\textrm{wt}}$, (B) $\bm{X}''_{\textrm{min}}$, (C)
  $\bm{X}''_{\textrm{bio}}$ and (D) $\bm{X}''_{\textrm{noP}}$. A
  parasegment is represented by four rounded rectangles, one for each
  cell, where the anterior cell is at the top, and posterior at the
  bottom. Since enput power is computed for every node in each of its
  two possible states, every cell rectangle has two rows of
  circles. The bottom row (marked on the sides with a white circle on
  the outside) corresponds to enput power of the nodes when
  \emph{off}, while the top row is the enput power when the same nodes
  are \emph{on} (marked on the sides with a dark circle). Each circle
  inside a cell's rectangle corresponds to the power of a given enput
  in the corresponding subset of MCs identified by the letters A to
  D. Full power is highlighted in red, other values in blue and
  scaled, while null power is depicted using small grey circles. Full
  power occurs only for enputs that are present in every MC (and
  configurations) of the respective set, whereas null power identifies
  nodes that are never enputs in any MC -- always irrelevant for the
  respective dynamical behaviour.}
\label{fig:epsilon}
\end{figure}

We studied the wild-type attractor basin of the spatial SPN model
using the four MC sets of interest: $\bm{X}''_{\textrm{wt}}$,
$\bm{X}''_{\textrm{min}}$, $\bm{X}''_{\textrm{bio}}$, and
$\bm{X}''_{\textrm{noP}}$ (see Minimal configurations subsection
above) focusing on the power of literal enputs only.
It is also possible to compute the enput power of group-invariant
enputs.
For example, the two-symbol MC $\bm{x}''_1$ in Figure
\ref{fig:min_two_sym}, has one of its four group-invariant enputs
defined by $ci = 1 \lor CI = 1$.
The power of this enput would tally those MCs in which this condition
holds.
Nonetheless, here we only measure the power of literal enputs and
present the study of the power of group-invariant enputs elsewhere.
The enput power computed for these four sets is depicted in
Figure \ref{fig:epsilon}, where the output nodes PH and SMO
are omitted because they are never input variables to any node in the
SPN model, and therefore have null power.
For the discussion of these results, it is useful to compare them to
the known initial condition, $\bm{x}_{\textrm{ini}}$ depicted in
Figure \ref{fig:spn_parasegment}, and the wild-type attractor,
$\mathcal{A}_{\textrm{wt}}$ depicted in Figure
\ref{fig:spatial_attractors} (a).

\textbf{Enput power in} $\bm{X}''_{\textrm{wt}}$ (see Figure \ref{fig:epsilon}A).
The enputs with \emph{full power} ($\epsilon=1$) are: SLP$_{1,2} = 0$,
SLP$_{3,4} = 1, hh_{2,4} = 0$ and $ptc_1 = 0$.
This is not entirely surprising since all of these genes and proteins
are specified as such in both $\bm{x}_{\textrm{ini}}$ and
$\mathcal{A}_{\textrm{wt}}$.
However, these values show that these enputs must remain in these
states in the entire (sampled) wild-type basin of attraction.
In other words, these enputs are \emph{critical controllers} of
the dynamics to the wild-type attractor.
Indeed, the wild-type is not \emph{robust} to changes in these enputs,
which are likely to steer the dynamics to other attractors, as
discussed further in the next section.
Therefore, the spatial SPN model appears to be unable to recover the
dynamic trajectory to the wild-type attractor when either the hedgehog
gene is expressed in cells two and four; or the patched gene is
expressed in the anterior cell, as well when the initial expression
pattern of SLP determined upstream by the pair-rule gene family is
disrupted in any way.
There are also enputs with \emph{high power} to control wild-type
behaviour: $wg_{1,3} = \textrm{WG}_{1,3} = 0$, $en_{1} = 1$, PTC$_1 = 0$,
$en_{2,4}=0$, $ptc_3 = 1$, CI$_3 = 0$ and CIR$_3 = 1$.
Again, these are the states of these genes and proteins in the known
initial configuration of the SPN $\bm{x}_{\textrm{ini}}$, and most of
them, except for $ptc_3 = 1$, CI$_3 = 0$ and CIR$_3 = 1$ correspond to their
final states in $\mathcal{A}_{\textrm{wt}} $.

In Figure \ref{fig:epsilon}A every node in the SPN --
except the omitted nodes PH and SMO -- appear as an enput, in at least
one Boolean state, in many cases with very low values of $\epsilon$.
Thus, while macro-level dynamics is significantly canalized (see
above), especially by SLP and the spatial signals for each cell,
control of wild-type can derive from alternative strategies,
whereby every node can act as an enput in some context.
Nonetheless, most nodes ultimately do not observe much power to
control wild-type behaviour, thus interventions to disturb wild-type
behaviour are most effective via the few more powerful controllers (see
also next section).

We can also compare the enput power computed for $\bm{X}''_{\textrm{wt}}$ (Figure \ref{fig:epsilon}A), with the
two-symbol MCs $\bm{x}''_1$ and $\bm{x}''_2$ in Figure
\ref{fig:min_two_sym}.
These two MCs redescribe a significant portion of the wild-type
attractor basin -- $20\%$ of our lower bound count of this
basin.
Because they only appear in $\bm{X}''_{\textrm{wt}}$ and not in
any of the other MC sets we studied, the portion of the wild-type
attractor basin they redescribe is unique to $\bm{X}_{\textrm{wt}}$,
and can be analysed via $\bm{x}''_1$ and $\bm{x}''_2$.
Most of the literal enputs specified in $\bm{x}''_1$ and $\bm{x}''_2$ have
high power in $\bm{X}''_{\textrm{wt}}$, except for WG$_2 = wg_4 =
\textrm{CIR}_{1,2,4} = 1$, which are enputs in these two-symbol MCs that
have low power.
Conversely, there are literal enputs with high-power in
$\bm{X}''_{\textrm{wt}}$ that are not enputs in these two-symbol MCs:
EN$_{2,4} = 0$ and PTC$_1=0$.
A key distinguishing feature of $\bm{x}''_1$ and $\bm{x}''_2$
is the expression of CIR across the entire parasegment as well as of
the wingless protein in the second cell, both of which are different from
the trajectory between the known initial condition of the SPN and the
wild-type attractor.
Therefore, $\bm{x}''_1$ and $\bm{x}''_2$ redescribe a (large) portion of the attractor basin outside of the more commonly studied dynamical trajectories.

\textbf{Enput power in} $\bm{X}''_{\textrm{min}}$ (see Figure
\ref{fig:epsilon}B).
We found an unexpected expression of CIR$_2 = 1$ (now with full power)
as well as $wg_2 = \textrm{WG}_2 = 1$ (high power).
Other enputs whose expression is in opposition to both
$\bm{x}_{\textrm{ini}}$ and $\mathcal{A}_{\textrm{wt}}$ appear with
low power: HH$_{2,4} = 1$ and CIR$_{1} = 1$.
This again suggests that there is a substantial subset of the
wild-type attractor basin, controlled by these and other enputs,
distinct from the trajectory that results from the known (biologically
plausible) initial configuration.
We can also see that there is a significant number of nodes that do
not play the role of enput in any MC -- nodes with \emph{null power},
depicted as small grey circles -- as well as many more enputs with
full power.
$\bm{X}''_{\textrm{min}}$ redescribes wild-type dynamics with the
smallest number (23) of enputs; this set contains only 32 MCs out of
the 1731 in $\bm{X}''_{\textrm{wt}}$.
However, these are the most macro-canalizing MCs that guarantee
convergence to wild-type. Indeed, because of their parsimony, they
redescribe a very large subset of the wild-type attractor basin with
at least 1.6 times more configurations than what was previously
estimated for this basin (see above).
Therefore, $\bm{X}''_{\textrm{min}}$ provides a solid baseline for the
understanding of control in the wild-type attractor basin.
This means that the genes and proteins with full power in this set are
critical controllers of wild-type behaviour.

\textbf{Enput power in} $\bm{X}''_{\textrm{bio}}$ (see Figure
\ref{fig:epsilon}C).
Because this MC set only redescribes configurations in the dynamic
trajectory from $\bm{x}_{\textrm{ini}}$ to $\mathcal{A}_{\textrm{wt}}
$, the transient dynamics observed in $\bm{X}''_{\textrm{wt}}$ and
$\bm{X}''_{\textrm{min}}$, e.g. $wg_2 = 1$ and CIR$_2 = 1$,
disappear.
There are, however, other enputs with full power: $wg_{1,3} =
\textrm{WG}_{1,3} = 0$, $en_{2,4} = \textrm{EN}_{2,4} = 0$, $ptc_1 =
\textrm{PTC}_1 = 0$.
These critical enputs are particularly important for restricting
analysis to a better-known portion of the wild-type attractor basin,
for which the model was especially built.

\textbf{Enput power in} $\bm{X}''_{\textrm{noP}}$ (see Figure
\ref{fig:epsilon}D).
This set of MCs is useful to understand the beginning of the segment
polarity regulatory dynamics, with no proteins expressed. The set of critical genes that must be
expressed (\emph{on}) are $ptc_3$ and $wg_4$, which appear with full power; moreover, $en_1 = hh_1 = ptc_2 = ci_2 = 1$ appear with high power.
As shown in the figure, most other enputs with full or high power
correspond to genes and proteins that must be inhibited (\emph{off}), except,
of course, SLP$_{3,4}$ that are assumed to be always \emph{on} in the
SPN model.

We compared these results with previous work on identifying
critical nodes in the SPN model.
Chaves et al. \cite{Chaves:2005fk} deduced, from the model's logic,
minimal `pre-patterns' for the initial configuration of the SPN that
guarantee convergence to wild-type attractor.
More specifically, two necessary conditions and one sufficient
condition were deduced, which we now contrast with the enput power
analysis.

The \textbf{first necessary condition} for convergence to the
wild-type attractor is: $ptc_3 = 1$, assuming that all proteins are
unexpressed (\emph{off}) initially, and the sloppy pair gene rule is
maintained constant (i.e. SLP$_{1,2} = 0 \; \land $ SLP$_{3,4} = 1$.)
Of the MC sets we analysed, only $\bm{X}''_{\textrm{noP}}$ obeys the
(biologically plausible) assumptions for this necessary condition. As we
can see in Figure \ref{fig:epsilon}D, the enput $ptc_3 = 1$ has full
power on this MC set, which confirms this previous theoretical
result. However, since every enput with full power is a
necessary condition for the set of configurations described by its MC
set, we can derive
other necessary conditions for this set of configurations (with the same assumptions), such as
$ptc_1 = 0$, $wg_3 = 0$, or $wg_4 = 1$ (see below). We can also see
that not all assumptions for the first necessary condition are
necessary; while the sloppy pair rule appears as four enputs with full
power, not all proteins are required to be unexpressed: the expression
of HH is irrelevant in every cell of the parasegment, as is the expression of PTC$_{2,3}$,
WG$_{2,4}$, CIA$_{4}$, and CIR$_{1,2,3}$. Moreover, the enput power
analysis allows us to identify `degrees of necessity'; some enputs may
not be necessary, but almost always necessary. This is the case of the
expression of $en_{1}$, which has high
power in $\bm{X}''_{\textrm{noP}}$, but is not a necessary condition as a few MCs can guarantee
convergence to wild-type with $en_{1} = 0$ (which also appears as
enput with low power).
Naturally, if we relax the assumptions for condition $ptc_3 = 1$, it
may no longer be a necessary condition. This can be see when we look
at the enput power analysis of the entire (sampled) wild-type basin
$\bm{X}''_{\textrm{wt}}$ (Figure \ref{fig:epsilon}A) or the smaller
$\bm{X}''_{\textrm{bio}}$ (Figure \ref{fig:epsilon}C). In these cases,
which still preserve the sloppy pair rule assumption, $ptc_3 = 1$ is
no longer an enput with full power. This means that, according to this
model, if some proteins are expressed initially, $ptc_3 = 1$ is no
longer a necessary condition.
Interestingly, we found that in the most macro-canalizing subset of
the attractor basin, $\bm{X}''_{\textrm{min}}$ (Figure
\ref{fig:epsilon}B) -- which assumes the sloppy pair rule constraint
but is not constrained to initially unexpressed proteins -- $ptc_3 =
1$ does appear as an enput with full power again. This means that in
the most parsimonious means to control convergence to wild-type
attractor, $ptc_3 = 1$ is a necessary condition too. It is noteworthy
that in this case, not only can some proteins be expressed, but the
expression of CIR$_{2}$ is also a necessary condition (enput with full
power).

\begin{figure}
\includegraphics{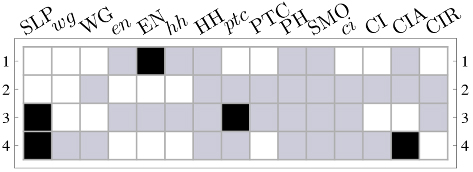}
\caption{{\bf A MC not requiring $wg_4=1 \lor en_1 =1 \lor ci_4 =
    1$ in wild-type attractor basin.} When proteins are allowed to be
  expressed initially, the second necessary condition, reported
  in \cite{Chaves:2005fk}, ceases to be a necessary condition, as
  discussed in the main text; in the MC shown, $wg_4, en_1$ and $ci_4$
  can be in any state and the network still converges to the wild-type
  attractor.}
\label{fig:chavessecond}
\end{figure}

The \textbf{second necessary condition} for convergence to the
wild-type attractor is: $wg_4 =1 \vee en_1 =1 \vee ci_4 =1$, assuming
that all proteins are unexpressed (\emph{off}) initially, and the
sloppy pair gene rule is maintained constant (i.e. SLP$_{1,2} = 0 \; \land $ SLP$_{3,4} = 1$) \cite{Chaves:2005fk}.
Again, only $\bm{X}''_{\textrm{noP}}$ obeys the (biologically likely)
assumptions for this necessary condition.
As we can see in Figure \ref{fig:epsilon}D, the enput $wg_4 =1$ has
full power, therefore it is a necessary condition.
However, the enput $en_1 =1$ has high power, and the enput $ci_4 =1$
has no power.
This means that they are not necessary, though $en_1 = 1$ is most
often needed.
These results suggest that this necessary condition could be shortened
to $wg_4 =1$, because in our sampling of the wild-type attractor
basin, in the subset meeting the assumptions of the condition, we did
not find a single configuration where $wg_4 =0$.
Even though our stochastic search was very large, it is possible that
there may be configurations, with no proteins expressed, where $wg_4 =
0 \; \land (en_1 = 1 \vee ci_4
=1 )$, thus maintaining the original necessary condition.
However, our enput power analysis gives a more realistic and nuanced
picture of control in the SPN model under the same assumptions. While
the necessary condition may be $wg_4 =1 \vee en_1 =1 \vee ci_4 =1$,
the individual enputs have strikingly different power in controlling
for wild-type behaviour: $ci_4 =1$ was never needed (no power), $en_1
=1$ has high power, and $wg_4 =1$ has full power.
Naturally, if we relax the assumptions for this condition, it may no
longer be a necessary condition. For instance, if we allow proteins to
be expressed initially (still preserving the sloppy pair constraint),
we can find MCs that redescribe configurations where $wg_4 = en_1 =
ci_4 = 0$.
We found 171 MCs in $\bm{X}''_{\textrm{wt}}$ (available in
\emph{Supporting data S14} where this condition is
not necessary, one of them depicted in Figure \ref{fig:chavessecond}.

The \textbf{sufficient condition} for convergence to the wild-type
attractor is: $wg_4 = 1 \; \land \; ptc_3 = 1$, assuming that the sloppy pair
gene rule is maintained constant (i.e. SLP$_{1,2} = 0 \; \land $
SLP$_{3,4} = 1$).
A variation of this sufficient condition assumes instead (maintaining
the sloppy pair gene rule): $wg_4 = 1 \; \land$ PTC$_{3}=1$
In their analysis, Chaves et al. \cite{Chaves:2005fk} assume that all
proteins are unexpressed and that many other genes are initially
inhibited (\emph{off}).
Even though in Chaves et
  al. \cite{Chaves:2005fk} the initial condition itself only requires
  $ptc_{1}=ci_{1,3}=0$, the argument hinges on propositions and facts
  that require knowing the state of additional genes such as $en_{2} =
  wg_{3} = hh_{2,4}=0$.
While Chaves et al. \cite{Chaves:2005fk} concluded rightly from this
minimal pre-pattern, that convergence to the wild-type pattern has a
remarkable error correcting ability to expression \emph{delays} in all
other genes, the condition does not really describe robustness to
\emph{premature expression} of genes and proteins.
It is interesting to investigate sufficient conditions that do require
the states of most variables to be specified, giving us the ability to
study robustness to both delays and premature expression of chemical
species.
The MC schemata we obtained with our macro-level analysis allows us to
investigate such sufficient conditions directly.

\begin{figure}
\includegraphics{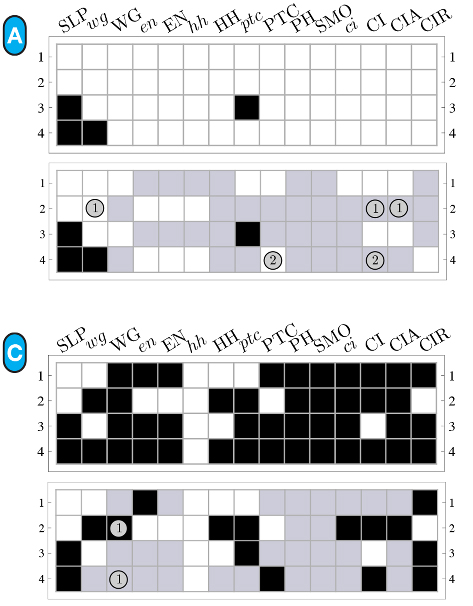}
\caption{{\bf `Extreme' configurations converging to wild-type in
    the SPN model.} (A) A configuration with the minimal number of
  nodes expressed that converges to wild-type, and its corresponding MC:
  32 nodes are irrelevant, 24 must be unexpressed (\emph{off}), and
  only 4 must be expressed (\emph{on}). (B) The opposite extreme
  condition where 16 genes and proteins are unexpressed and all other 44
  are expressed.}
\label{fig:extremes}
\end{figure}

We searched the entire MC
set $\bm{X}''_{\textrm{wt}}$ to retrieve the MCs with the \emph{fewest}
number of enputs specified as \emph{on}.
The 10 MCs (available in \emph{S11}) we retrieved contain only 26 literal enputs,
where in six MCs the two nodes in the sufficient condition above ($wg_4, ptc_3$), plus the
nodes from the sloppy pair rule (SLP$_{3,4}$) are \emph{on}, 24 are
\emph{off} and the remaining 32 are wildcards, and thus irrelevant.
In the remaining MCs, instead of $ptc_3 = 1$, we found PTC$_3 = 1$ to
be an enput. In those MCs $ptc_3 = \#$.
Converting all wildcards to \emph{off} in one of these MCs, confirms
the sufficient condition, as can be seen from Figure
\ref{fig:extremes}A, where SLP$_{3,4} = wg_4 = ptc_3 =
1$, and everything else is \emph{off}.
This can be seen as an `extreme' condition to wild-type attractor, with
a minimum set of genes expressed.
We also searched for the opposite extreme scenario, retrieving all MCs
with the largest number of \emph{on} nodes, that still converges to
the wild-type pattern (available in \emph{Supporting data S12}.  
By replacing all wildcards in such MCs to \emph{on}, we obtained the
configuration in which only 16 nodes must be inhibited (\emph{off}),
while the remaining 44 are expressed (\emph{on}), depicted in Figure
\ref{fig:extremes}B.
Interestingly, in this extreme configuration, $hh$ must remain
\emph{off} across the whole parasegment.

\subsection*{Robustness to enput disruption}
%\addcontentsline{toc}{subsection}{Robustness to enput disruption}

The power measure introduced in the previous subsection allows us to
predict critical nodes in controlling network dynamics to a pattern of
interest $\bm{\mathcal{P}}$.
A natural next step is to investigate what happens when the critical
controllers are actually disrupted.
We can disrupt an enput $e$ in an MC set with a variety of dynamic
regimes.
Here, we adopt the approach proposed by Helikar \emph{et al.}
\cite{Helikar:2008fk}, where a node of interest flips its state at
time $t$ with a probability $\zeta$, which can be seen to represent noise in
regulatory and signalling events, as well as the `concentration' of a
gene (its corresponding mRNA) or protein -- thus making it possible to use Boolean networks
to study continuous changes in concentration of biochemical systems
(see \cite{Helikar:2008fk}).

We start from an initial set of configurations of interest:
$\bm{X}^0$. This can be a single configuration, such as the known
initial configuration of the SPN $\bm{X}^0 \equiv
\{\bm{x}_{\textrm{ini}}\}$ (as in Figure \ref{fig:spn_parasegment}A), where the enput $e$ is in a specific (Boolean) value.
Next, we set the value of \emph{noise} parameter $\zeta$, which is the
probability that $e$ momentarily flips from its state in $\bm{X}^0$ at
time $t$.
This noise is applied at every time step of the simulated dynamics;
when a state-flip occurs at time $t$, the node returns to its original
state at $t+1$ when noise with probability $\zeta$ occurs again.
Noise is applied to $e$ from $t = 0$ to $t = m$.
At time step $t = m+1$ no more noise is applied to $e$ ($\zeta = 0$)
and the network is allowed to converge to an attractor.
This process is repeated for $M$ trials.
Finally, we record the proportions of the $M$ trials that converged to
different attractors.

Since in this paper we only computed enput power for literal enputs
(see previous subsection), we also only study literal enput
disruption. It is straightforward to disrupt group-invariant enputs;
for instance, the group-invariant enput defined by $ci = 1 \lor$ CI $=
1$ from the two-symbol MC $\bm{x}''_1$ in Figure
\ref{fig:min_two_sym}, can be perturbed by making $ci = 0 \land$ CI $=
0$. Nonetheless, for simplicity, we present the study of the
disruption of group-invariant enputs elsewhere.

The enput power analysis in the previous subsection, revealed that in
the wild-type attractor basin ($\bm{X}_{\textrm{wt}}$) of the spatial
SPN model there are the following critical nodes (or key controllers):
across the parasegment, SLP proteins must be inhibited in cells 1 and
2 (SLP$_{1,2} = 0$) and expressed in cells 3 and 4 (SLP$_{3,4} = 1$),
as determined by the pair-rule gene family; hedgehog genes (spatial
signals) in cells 2 and 4 must be inhibited ($hh_{2,4} = 0$); the
patched gene in the anterior cell must also be inhibited ($ptc_1 =
0$).
With the \emph{stochastic intervention} procedure just described, we
seek to answer two questions about these key controllers:
(1) how sensitive are they to varying degrees of stochastic noise? and
(2) which and how many other attractors become reachable when they are
disrupted?
In addition to the seven full power enputs, for comparison purposes,
we also test the low power enput CI$_4 = 0$.
In the original SPN model the states of SLP$_{1,2,3,4}$ are fixed (the
sloppy gene constraints).
Because these naturally become enputs with full power (see Figure
\ref{fig:epsilon}), it is relevant to include them in this study of
enput disruption.
However, by relaxing the fixed-state constraint on SLP$_{1,2,3,4}$, by
inducing stochastic noise, the dynamical landscape of the spatial SPN
model is enlarged from $2^{56}$ to $2^{60}$ configurations.
This means that more attractors than the ten identified for the SPN
Boolean model (depicted in Figure \ref{fig:spatial_attractors}) are
possible, and indeed found as explained below.

\begin{figure}
\includegraphics{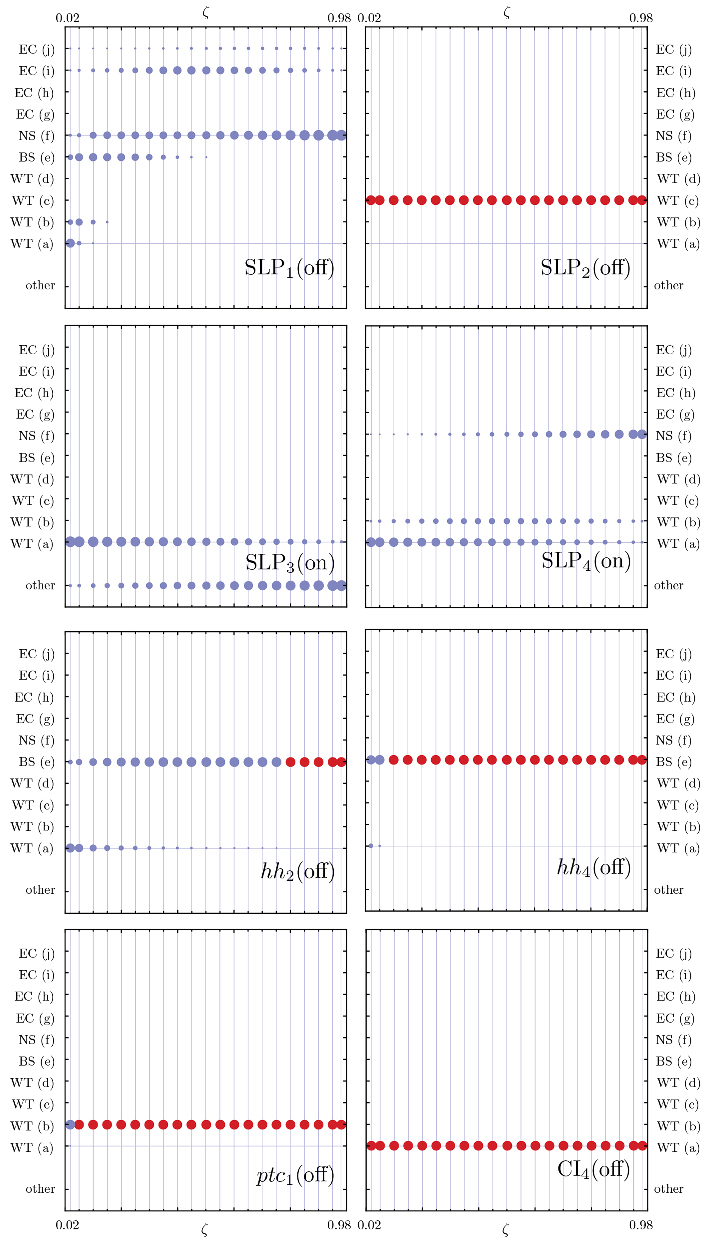}
\caption{{\bf Wild-type enput disruption in the SPN model.} Each
  coordinate $(x,y)$ in a given diagram (each corresponding to a tested
enput) contains a circle, depicting the proportion of trials
that converged to attractor $y$ when noise level $x$ was
used. Red circles mean that all trajectories tested converged to $y$.}
\label{fig:stochastic}
\end{figure}

We used $\bm{X}^0 \equiv \{\bm{x}_{\textrm{ini}}\}$ as the initial
state of the networks analysed via stochastic interventions, because
of its biological relevance.
The simulations where performed with the following parameters:
$\zeta \in [0.05,0.95]$, swept with $\Delta(\zeta) = 0.05$, plus extremum values $\zeta =0.02$ and $\zeta
=0.98$;
$m = 500$ steps; $M=10^4$.
The simulation results are shown in Figure \ref{fig:stochastic}.

The first striking result is that disruption of SLP$_1 = 0$ makes it
possible to drive the dynamics away from wild-type into one of five
other attractors (one of which a variant of wild-type).
For $\zeta > 0.15$ no further convergence to wild-type is observed, and at
$\zeta =0.05$ the proportion of trials that converged to wild-type was
already very small.
We also found phase transitions associated with the values of $\zeta$.
For $\zeta \le 0.15$  most trials converged to wild-type, wild-type
(ptc mutant), broad-stripes or no-segmentation, and a very small
proportion to two variants of the ectopic mutant.
When $\zeta = 0.15$ the proportion of trials converging to
broad-stripes reaches its peak, and decreases, so that no trial
converged to this mutant expression pattern for $\zeta \ge 0.55$.
Finally, for $\zeta \ge 0.55$ convergence to the ectopic variants
reaches its peak and decreases steadily but does not disappear, while
convergence to the no-segmentation mutant increases becoming almost
$100\%$ when $\zeta = 0.98$. We thus conclude that SLP$_1 = 0$ is a wild-type attractor enput which is very sensitive to noise.

In the case of SLP$_{3} = 1$, we observed convergence to an attractor
that is not any of the original ten attractors -- characterized by
having two engrailed bands in cells 1 and 3 (see \emph{Supporting text S5}).
The proportion of trials converging to wild-type and to the new attractor
decrease and increase respectively, reaching similar proportions when
$\zeta=0.5$.
When $\zeta=0.98$, almost every trial converged to the new
attractor.
We conclude that SLP$_{3} = 1$ is a wild-type attractor enput whose
robustness is proportional to noise.

Disruption of SLP$_4 = 1$ resulted in a behaviour
similar to SLP$_1$, but with fewer possible attractors reached.
As $\zeta$ is increased, fewer trials converge to wild-type and growing
proportions of trials converge to the wild-type $ptc$ mutant pattern
(reaching a peak at $\zeta =0.5$) and the no-segmentation mutant.
For more extreme values of $\zeta$, the majority of trials converged
to the no-segmentation mutant.
However, an important difference with respect to SLP$_1$ was observed:
for $\zeta \le 0.5$ the majority of trials converged to wild-type, and
convergence to this attractor is observed for the whole range of
$\zeta$.
Thus the wild-type phenotype in the SPN model is much more robust to
perturbations to the expression of SLP in the posterior cell (SLP$_4 =
1$), than to perturbations to its inhibition in the anterior cell
(SLP$_1 = 0$).

With the parameters chosen, the disruption of SLP$_2= 0$ leads to a
remarkable similar behaviour: any disruption (any amount of noise)
leads to the same wild-type variant attractor pattern with two
wingless stripes (c). Therefore, SLP$_2= 0$ is not robust at all --
though the resulting attractor is always the same and a variant of
wild-type.
In this case, convergence to a single attractor
for all values of $\zeta$ is the result of setting $m=500$ in our experiments.
When we lower the value of $m$ enough in our simulations, for low
values of $\zeta$, there are trials that are not perturbed and thus
maintain convergence to the wild-type attractor.
But any perturbation of SLP$_2= 0$ that occurs leads the dynamics to the wild-type variant.

Disruption of $hh_{2,4}=0$ increasingly drives
dynamics to the broad-stripes mutant.
However, disruption of $hh_2$ reveals greater robustness since a large
number of trials still converges to wild-type for $\zeta \le 0.15$,
and residual convergence to wild-type is observed up to $\zeta
= 0.75$.
In contrast, any disruption of $hh_4$ above $\zeta = 0.05$ leads to
the broad-stripes mutant, and even very small amounts of disruption
lead to a large proportion of mutants.
Similarly, disruption of $e \equiv ptc_1 = 0$ drives the dynamics
to one -- and the same -- of the wild-type variants.
Yet, when $\zeta=0.02$ there is a minute proportion of trajectories
that still converge to the wild-type attractor. Therefore, as
expected, the wild-type attractor in the SPN model is not very robust
to disruptions of the enputs with full power.
Finally, and in contrast, no disruption of low-power enput CI$_4= 0$
is capable of altering convergence to the wild-type attractor.

\section*{Discussion}
%\addcontentsline{toc}{section}{Discussion}

We introduced wildcard and two-symbol redescription as a means to
characterize the control logic of the automata used to model
networks of biochemical regulation and signalling.
We do this by generalizing the concept of \emph{canalization}, which becomes
synonymous with redundancy in the logic of automata. The two-symbol
schemata we propose capture two forms of logical redundancy, and
therefore of canalization: input redundancy and symmetry.
This allowed us to provide a straightforward way to \emph{quantify}
canalization of individual automata (micro-level), and to integrate the
entire canalizing logic of an automata network into the Dynamics
Canalization Map (DCM).
A great merit of the DCM is that it allows us to make inferences about
collective (macro-level) dynamics of networks from the micro-level canalizing logic
of individual automata -- with incomplete information.
This is important because even medium-sized automata models of
biochemical regulation lead to dynamical landscapes that are too large
to compute. In contrast, the DCM scales linearly with number of
automata -- and schema redescription, based on computation of prime
implicants -- is easy to compute for individual automata with the number
of inputs typically used in the literature.

With this methodology, we are thus providing a method to link micro- to
macro-level dynamics -- a crux of complexity.
Indeed, in this paper we showed how to uncover \emph{dynamical modularity}:
separable building blocks of macro-level dynamics. This an entirely
distinct concept from community structure in networks, and allows us
to study complex networks with node dynamics -- rather than just their connectivity
structure.
The identification of such modules in the dynamics of networks is
entirely novel and provides insight as to how the collective dynamics
of biochemical networks uses these building blocks to produce its
phenotypic behaviour -- towards the goal of explaining how biochemical
networks `compute'.

By basing our methodology on the redescription of individual automata
(micro-level), we also avoid the scaling problems faced by previous
schemata approaches which focused solely on redescription of the
dynamical landscape (macro-level) of networks \cite{Willadsen:2007hc}.
By implementing the DCM as a threshold network, we show that we can
compute the dynamical behaviour of the original automata network from
information about the state of just a few network nodes (partial
information).
In its original formulation, the dynamic unfolding of an automata
network cannot be computed unless an initial state of all its nodes is
specified.
In turn, this allows us to search for minimal conditions (MCs) that
guarantee convergence to an attractor of interest. 
Not only are MCs important to understand how to \emph{control} complex
network dynamics, but they also allow us to \emph{quantify macro-level
  canalization} therein.
From this, we get a measurable understanding of the robustness of
attractors of interest -- the greater the canalization, the greater
the robustness to random perturbations -- and, conversely, the
identification of \emph{critical node-states} (enputs) in the network
dynamics to those attractors.
We provided a measure of the capacity of these critical nodes to
control convergence to an attractor of interest (enput power), and
studied their robustness to disruptions.
By quantifying the ability of individual nodes to control attractor
behaviour, we can obtain a testable understanding of macro-level
canalization in the analysed biochemical network. 
Indeed, we can uncover how robust phenotypic traits are
(e.g. robustness of the wild-type attractor), and which critical nodes
must be acted upon in order to disrupt phenotypic behaviour.

We exemplified our methodology with the well-known segment polarity
network model (in both the single-cell and the spatial versions).
Because this model has been extensively studied, we use it to show
that our analysis does not contradict any previous findings. However,
our analysis also allowed us to gain new knowledge about its
behaviour. From a better understanding of the size of its wild-type
attractor basin (larger than previously thought) to uncovering
new minimal conditions and critical nodes that control wild-type
behaviour. We also fully quantified micro- and macro-level
canalization in the model, and provided a complete map of its
canalization logic including dynamical modularity.
Naturally, our results pertain to this model; we do not claim that our
results characterize the real Drosophila segment polarity gene
network. However, our results, should they be found to deviate from
organism studies, can certainly be used to improve the current model,
and thus improve our understanding of Drosophila development.
Thus a key use of our methodology in systems biology should be to help
improve modelling accuracy.
With the methodology now tested on this model, in subsequent work we
will apply it to several automata network models of biochemical
regulation and signalling available in the systems biology literature.

The pathway modules we derived by inspection of the DCM for the
segment polarity network revealed a number of properties of complex
networks dynamics that deserve further study.
For instance, the dynamical sequence that occurs once each such module
is activated is independent of the temporal update scheme utilized. 
Therefore, if the dynamics of a network is captured exclusively by
such modules, its intra-module behaviour will be similar for both
synchronous and asynchronous updating -- denoting a particular form of
robustness to timing.
We will explore this property in future work, but as we showed here,
the dynamics of the single-cell version of the SPN model is very
(though not fully) controlled by only two pathway modules. This
explains why its dynamical behaviour is quite robust to timing events
as previously reported \cite{Chaves:2005fk}.

Research in cellular processes has provided a huge amount of genomic,
proteomic, and metabolomics data used to characterize networks of
biochemical reactions. All this information opens the possibility of
understanding complex regulation of intra- and inter-cellular
processes in time and space. However, this possibility is not yet
realized because we do not understand the dynamical constraints that
arise at the phenome (macro-) level from micro-level interactions.
One essential step towards reaching these ambitious goals is to
identify and understand the loci of control in the dynamics of complex
networks that make up living cells.
Towards this goal, we developed the new methodology presented in this paper.
Our methodology is applicable to any complex network that can be
modelled using binary state automata -- and easily extensible to
multiple-state automata.
We currently focus only on biochemical regulation with the goal of
understanding the possible mechanisms of collective information
processing that may be at work in orchestrating cellular activity.

% Do NOT remove this, even if you are not including acknowledgments
\section*{Acknowledgements}

%This work was supported by Funda\c{c}\~ao para a Ci\^encia e a Tecnologia (Portugal) grant \emph{PTDC/EIA-CCO/114108/2009}.

We thank the FLAD Computational Biology Collaboratorium at the
Gulbenkian Institute of Science (Portugal) for hosting and providing facilities
used for this research. We also thank Indiana University for providing
access to its computing facilities. Finally, we are very grateful for
the generous and constructive comments we received from reviewers.

\cleardoublepage
%\section*{References}
% The bibtex filename
\bibliography{MMPRefs}

\cleardoublepage
\section*{Tables}

\begin{table}
{\small
\[
\begin{array}{|l|l|l|}
\hline
\text{\emph{Index}} & \text{\emph{Node}} & \text{\emph{State-Transition Function}} \\
\hline
1 & \textrm{SLP}_i^{t+1} & \leftarrow 0 \text{ if } i = 1 \lor i = 2; 1 \text{ if } i = 3 \lor i = 4;  \\
2 & wg_i^{t+1} & \leftarrow (\textrm{CIA}_i^t \land \textrm{SLP}_i^t \land \neg \textrm{CIR}_i^t)\lor (wg_i^t \land (\textrm{CIA}_i^t \lor \textrm{SLP}_i^t)\land \neg \textrm{CIR}_i^t) \\
3 & \textrm{WG}_i^{t+1} & \leftarrow wg_i^t \\
4 & en_i^{t+1} & \leftarrow ({\textrm{WG}_{i-1}^t} \lor {\textrm{WG}_{i+1}^t}) \land \neg \textrm{SLP}_i^t \\
5 & \textrm{EN}_i^{t+1} &  \leftarrow en_i^t \\
6 & hh_i^{t+1} &  \leftarrow \textrm{EN}_i^t \land \neg \textrm{CIR}_i^t \\
7 & \textrm{HH}_i^{t+1} &  \leftarrow hh_i^t \\
8 & ptc_i^{t+1} & \leftarrow \textrm{CIA}_i^t \land \neg \textrm{EN}_i^t \land \neg \textrm{CIR}_i^t \\
9 & \textrm{PTC}_i^{t+1} & \leftarrow ptc_i^t \lor (\textrm{PTC}_i^t \land \neg {\textrm{HH}_{i-1}^t} \land \neg {\textrm{HH}_{i-1}^t}) \\
10 & \textrm{PH}_i^{t} & \leftarrow \textrm{PTC}_i^t \land ({\textrm{HH}_{i-1}^t} \lor {\textrm{HH}_{i+1}^t}) \\
11 & \textrm{SMO}_i^{t} & \leftarrow \neg \textrm{PTC}_i^t \lor  ({\textrm{HH}_{i-1}^t} \lor {\textrm{HH}_{i+1}^t}) \\
12 & ci_i^{t+1} & \leftarrow \neg \textrm{EN}_i^t \\
13 & \textrm{CI}_i^{t+1} & \leftarrow ci_i^t \\
14 & \textrm{CIA}_i^{t+1} & \leftarrow \textrm{CI}_i^t \land (\neg \textrm{PTC}_i^t \lor  {hh_{i-1}^t} \lor {hh_{i+1}^t}  \lor {\textrm{HH}_{i-1}^t} \lor {\textrm{HH}_{i+1}^t}) \\
15 & \textrm{CIR}_i^{t+1} & \leftarrow \textrm{CI}_i^t \land \textrm{PTC}_i^t \land  \neg {hh_{i-1}^t} \land \neg {hh_{i+1}^t}  \land \neg  {\textrm{HH}_{i-1}^t} \land \neg  {\textrm{HH}_{i+1}^t} \\
\hline
\end{array}
\]
}
\caption{Boolean logic formulae representing the state-transition
  functions for each node in the SPN (four-cell parasegment)
  model. The subscript represents the cell index; the superscript
  represents time. Note that every node has a numerical index assigned
  to it, which we use for easy referral throughout the present
  work. The extra-cellular nodes, $hh, \textrm{HH}$ and $\textrm{WG}$
  in adjacent cells are indexed as follows:  16 to 21 denote  $
  hh_{i-1}$, $ hh_{i+1}$, $ \textrm{HH}_{i-1}$, $\textrm{HH}_{i+1}$,
  $\textrm{WG}_{i-1}$ and $\textrm{WG}_{i+1}$ in this order.}
\label{tab:SPN_Logic_Table}
\end{table}

\begin{table}

\begin{tabular}{|c|c|c|}
\hline
 & {\bf s-units} & {\bf t-units} \\
\hline
{\bf incoming fibres} & one or more & one or more\\
\hline
{\bf outgoing fibres} & one per schema of which is enput & one for the
transition it causes\\
\hline
{\bf branching out} & yes & no\\
\hline
{\bf fusing in} & no & yes\\
\hline
\end{tabular}
\caption{{\bf Connectivity rules in canalizing maps}}
\label{tab:TN_Rules_Wiring}
\end{table}

\begin{table}
  {\small
\begin{tabular}{|c|c|c|c|c|c|c|}
\hline
 {\bf MC set} & $|\bm{X}''|$ & $e$ (min) & $e$ (max) &$
 n_\textrm{e}$ & $n_\textrm{r}$ & $n_\textrm{s}$ \\
\hline
 $\bm{X}'_{\textrm{wt}}$ &  1745 & 23 & 33 &  $24.01 \pm 0.08$ & 35.99 $\pm 0.17$ & $0.98 \pm 0.03$ \\
\hline
$ \bm{X}'_{\textrm{min}}$ & 32 & 23 & 23 & $23  \pm 0$ & 37 $\pm 0$& 0 \\
\hline
 $\bm{X}'_{\textrm{bio}} $&  90 & 25  & 28 & $25.75 \pm  0.11$ &
 34.25 $\pm  0.11$ & 0 \\
\hline
$ \bm{X}'_{\textrm{noP}}$ &  24 & 26 & 30 & $26.2 \pm 0.04$ & 34.8 $\pm 0.04$ & 0 \\
\hline
\end{tabular}
}
  \caption{{\bf Macro-level canalization in the wildcard MC sets
      converging to wild-type in the SPN.} The table lists for every
    set of MCs reported in the main text: cardinality, minimum number of enputs, maximum number of enputs, estimated canalization. Canalization measures were obtained, for each MC set, from $10$ independent samples of $10^4$ configurations, thus $|\bm{\hat{X}}| = 10^5$. Values shown refer to the mean plus 95\% confidence intervals for the 10 independent measurements.}
\label{tab:MC_Summary}
\end{table}

\begin{table}

{\small
\begin{tabular}{|c|c|c|c|c|c|c|c|c|c|}
\hline
 {\bf MC set} & $n^*_\textrm{e}$ & $n^*_\textrm{r}$ & $n^*_\textrm{s}$\\
\hline
 $\bm{X}'_{\textrm{wt}}$ & 0.4 $\pm 0.001$ & 0.6 $\pm 0.001$ & 0.016 $\pm 0.002$\\
\hline
$ \bm{X}'_{\textrm{min}}$ & 0.38 & 0.62 & 0\\
\hline
 $\bm{X}'_{\textrm{bio}} $& 0.43 $\pm 0.001$ & 0.57 $\pm 0.001$& 0\\
\hline
$ \bm{X}'_{\textrm{noP}}$  & 0.436 $\pm 0.0007$& 0.564 $\pm 0.0007$ & 0\\
\hline
\end{tabular}
}
 \caption{{\bf Macro-level canalization in the wildcard MC sets
      converging to wild-type in the SPN.} The table lists the relative canalization measures for every
    set of MCs reported in the main text. Canalization measures were obtained, for each MC set, from $10$ independent samples of $10^4$ configurations, thus $|\bm{\hat{X}}| = 10^5$. Values shown refer to the mean plus 95\% confidence intervals for the 10 independent measurements.
    }
\label{tab:MC_Summary2}
\end{table}

\cleardoublepage
\section*{Figure Legends}

% FIGURE 1

% FIGURE 2

% FIGURE 3

% FIGURE 4

% FIGURE 5

% FIGURE 6

% FIGURE 7

% FIGURE 8

% FIGURE 9

% FIGURE 10

% FIGURE 11

% FIGURE 12

% FIGURE 13

% FIGURE 14

% FIGURE 15

% FIGURE 16

% FIGURE 17

% FIGURE 18

\cleardoublepage
% FIGURE 19

% FIGURE 20

% FIGURE 21

\cleardoublepage
\section*{Supporting Information Legends}

Supporting text {\bf S1} Glossary and mathematical notation. \\
Supporting text {\bf S2} Details about the computation of wildcard and
two-symbol schemata.\\
Supporting text {\bf S3} Details about the conversion of schemata into
a single threshold network.\\
Supporting text {\bf S4} Algorithms for the computation of minimal configurations. \\
Supporting text {\bf S5} Further details concerning the minimal
configurations found for the segment polarity network model. \\
Supporting text {\bf S6} Basic notions of the inclusion/exclusion principle. \\
Supporting data {\bf S7} (.csv format) Minimal configurations for the segment
polarity network model obtained from biologically-plausible seed configurations. \\
Supporting data {\bf S8} (.csv format)  Entire set of minimal configurations obtained
for the segment polarity network model. \\
Supporting data {\bf S9} (.csv format)  Minimal configurations for the segment
polarity network where no protein is \emph{on}. \\
Supporting data {\bf S10} (.csv format)  Minimal configurations for the segment
polarity network with the smallest number of nodes that need to be
specified in a Boolean state. \\
Supporting data {\bf S11} (.csv format) Minimal configurations for the segment
polarity network with the fewest number of  \emph{on} nodes \\
Supporting data {\bf S12} (.csv format) Minimal configurations for the segment
polarity network with the largest number of  \emph{on} nodes \\
Supporting data {\bf S13} (.csv format) (Wildcard) minimal configurations for the segment
polarity network that were redescribed as two-symbol schemata \\
Supporting data {\bf S14} (.csv format) Minimal configurations for the segment
polarity network that do not satisfy $wg_4 =1 \vee en_1 =1 \vee ci_4 =1$  \\

\end{document}